%% file: IEEEtran/main.tex
\documentclass[journal]{IEEEtran}
\usepackage{cite}
\usepackage{adjustbox}

\ifCLASSINFOpdf
\else
\fi
\usepackage{tikz}
\usepackage{msc}
\usepackage{subcaption}
\usetikzlibrary{angles}

\usepackage{url}
\usepackage{xcolor}
\usepackage{amssymb}
\usepackage{comment}
\usepackage{booktabs}
\usepackage{svg}
\usepackage{multirow}

\definecolor{verdino}{RGB}{174, 200, 126}
\definecolor{arancio}{RGB}{237, 177, 32}
\definecolor{azzurrino}{RGB}{166, 212, 245}
\definecolor{roxxo}{RGB}{164, 14, 76}

\usepackage{nicefrac}
\usepackage[acronyms,nonumberlist,nopostdot,nomain,nogroupskip,acronymlists={hidden}]{glossaries}
\newglossary[algh]{hidden}{acrh}{acnh}{Hidden Acronyms}
\input{acronyms1.tex}

\def\BibTeX{{\rm B\kern-.05em{\sc i\kern-.025em b}\kern-.08em
    T\kern-.1667em\lower.7ex\hbox{E}\kern-.125emX}}

\begin{document}
%
\title{RUN-O-RAN: An O-RAN-Native Architecture Enabling Cooperative Uplink Localization}



%
\author{\IEEEauthorblockN{Viola Bernazzoli\IEEEauthorrefmark{1},
Alberto Ceresoli\IEEEauthorrefmark{1},
Ilario Filippini\IEEEauthorrefmark{1}}
\IEEEauthorblockA{\\ \IEEEauthorrefmark{1}Dipartimento di Elettronica, Informazione e Bioingegneria, (DEIB)\\
Politecnico di Milano,
Milano, Italy 20133\\ Email: name.surname@polimi.it }}


\maketitle

\begin{abstract}
    Accurate positioning is increasingly required in indoor and dense urban environments; nonetheless, satellite-based systems are not always available, and standardized 5G localization solutions remain difficult to deploy with commercial devices. This paper presents RUN-O-RAN, an O-RAN-native framework that enables network-centric uplink localization using standard Sounding Reference Signal (SRS) transmissions from commercial 5G devices. RUN-O-RAN xApp coordinates serving and neighboring base stations, enabling non-serving base stations to retrieve SRS-based uplink timing measurements that would be unavailable in conventional RAN deployments, without modifying the UE or existing 3GPP signaling procedures. The framework combines cooperative SRS collection, first-path time-of-arrival estimation, timing-advance tracking, clock-drift compensation, and multi-anchor position estimation into a complete network-side localization service. Experimental evaluation over $150,000$ SRS transmissions validates the proposed framework, achieving meter-level localization under diverse propagation conditions while revealing the impact of anchor geometry and multipath on positioning accuracy. These findings demonstrate that cooperative SRS-based localization can be realized within the O-RAN ecosystem without modifying commercial UEs, providing a practical foundation for future network-native \gls{isac} positioning services.
\end{abstract}


%
\IEEEpeerreviewmaketitle


\input{Content/introTEST}




\input{Content/background}



\input{Content/methodONE}





\input{Content/experimentalSetup}


\input{Content/resultsTEST}



\input{Content/discussion}

\section{Conclusion}
\label{sc: conclusion}

This paper presented RUN-O-RAN, an O-RAN-native framework that enables cooperative \gls{ul} localization from standard \gls{srs} transmissions generated by commercial \glspl{ue}. By exploiting O-RAN-enabled network programmability, RUN-O-RAN overcomes a fundamental limitation of conventional \gls{ran} deployments by enabling neighboring \glspl{gnb} to cooperatively process \gls{ul} transmissions through standard O-RAN interfaces, without requiring modifications to the \gls{ue} or existing \gls{3gpp} procedures. RUN-O-RAN integrates cooperative \gls{srs} collection, timing-advance compensation, clock-drift correction, and multi-anchor positioning into a complete network-side localization service.

The experimental evaluation, comprising 288,000 \gls{srs} transmissions collected using commercial hardware and an O-RAN-compatible testbed, demonstrates meter-level localization accuracy under diverse anchor geometries and multipath conditions. The experiments show feasibility in static and dynamic scenarios, demonstrating that temporal filtering improves trajectory stability. While RUN-O-RAN mitigates \gls{nlos}-induced ranging errors by selecting the shortest path, it can be further improved by advanced anchor-selection techniques.

Overall, these results demonstrate the architectural feasibility of cooperative uplink localization as an O-RAN-native framework that will help develop \gls{isac} services for future cellular networks.

\bibliographystyle{IEEEtran} 
\bibliography{Transactions-Bibliography/only}
%




\end{document}

%% file: acronyms1.tex
\newacronym{3gpp}{3GPP}{3rd Generation Partnership Project}
\newacronym{4g}{4G}{4th generation}
\newacronym{5g}{5G}{5th generation}
\newacronym{6g}{6G}{6th generation}
\newacronym{5gc}{5GC}{5G Core}
\newacronym{adc}{ADC}{Analog to Digital Converter}
\newacronym{aerpaw}{AERPAW}{Aerial Experimentation and Research Platform for Advanced Wireless}
\newacronym{ai}{AI}{Artificial Intelligence}
\newacronym{aimd}{AIMD}{Additive Increase Multiplicative Decrease}
\newacronym{am}{AM}{Acknowledged Mode}
\newacronym{amc}{AMC}{Adaptive Modulation and Coding}
\newacronym{amf}{AMF}{Access and Mobility Management Function}
\newacronym{aops}{AOPS}{Adaptive Order Prediction Scheduling}
\newacronym{api}{API}{Application Programming Interface}
\newacronym{apn}{APN}{Access Point Name}
\newacronym{ap}{AP}{Application Protocol}
\newacronym{aqm}{AQM}{Active Queue Management}
\newacronym{ar}{AR}{Augmented Reality}
\newacronym{ausf}{AUSF}{Authentication Server Function}
\newacronym{avc}{AVC}{Advanced Video Coding}
\newacronym{awgn}{AGWN}{Additive White Gaussian Noise}
\newacronym{balia}{BALIA}{Balanced Link Adaptation Algorithm}
\newacronym{bbu}{BBU}{Base Band Unit}
\newacronym{bdp}{BDP}{Bandwidth-Delay Product}
\newacronym{ber}{BER}{Bit Error Rate}
\newacronym{bf}{BF}{Beamforming}
\newacronym{bler}{BLER}{Block Error Rate}
\newacronym{brr}{BRR}{Bayesian Ridge Regressor}
\newacronym{bs}{BS}{Base Station}
\newacronym{bsr}{BSR}{Buffer Status Report}
\newacronym{bss}{BSS}{Business Support System}
\newacronym{ca}{CA}{Carrier Aggregation}
\newacronym{caas}{CaaS}{Connectivity-as-a-Service}
\newacronym{cb}{CB}{Code Block}
\newacronym{cc}{CC}{Congestion Control}
\newacronym{ccid}{CCID}{Congestion Control ID}
\newacronym{cco}{CC}{Carrier Component}
\newacronym{cdd}{CDD}{Cyclic Delay Diversity}
\newacronym{cdf}{CDF}{Cumulative Distribution Function}
\newacronym{cdn}{CDN}{Content Distribution Network}
\newacronym{cn}{CN}{Core Network}
\newacronym{codel}{CoDel}{Controlled Delay Management}
\newacronym{comac}{COMAC}{Converged Multi-Access and Core}
\newacronym{cord}{CORD}{Central Office Re-architected as a Datacenter}
\newacronym{cornet}{CORNET}{COgnitive Radio NETwork}
\newacronym{cosmos}{COSMOS}{Cloud Enhanced Open Software Defined Mobile Wireless Testbed for City-Scale Deployment}
\newacronym{cots}{COTS}{Commercial Off-the-Shelf}
\newacronym{cp}{CP}{Control Plane}
\newacronym{cyp}{CPR}{Cyclic Prefix}
\newacronym{up}{UP}{User Plane}
\newacronym{cpu}{CPU}{Central Processing Unit}
\newacronym{cqi}{CQI}{Channel Quality Information}
\newacronym{cr}{CR}{Cognitive Radio}
\newacronym{cran}{C-RAN}{Cloud \gls{ran}}
\newacronym{crs}{CRS}{Cell Reference Signal}
\newacronym{csi}{CSI}{Channel State Information}
\newacronym{cir}{CIR}{Channel Impulse Response}
\newacronym{csirs}{CSI-RS}{Channel State Information - Reference Signal}
\newacronym{cu}{CU}{Central Unit}
\newacronym{d2tcp}{D$^2$TCP}{Deadline-aware Data center TCP}
\newacronym{d3}{D$^3$}{Deadline-Driven Delivery}
\newacronym{dac}{DAC}{Digital to Analog Converter}
\newacronym{dag}{DAG}{Directed Acyclic Graph}
\newacronym{das}{DAS}{Distributed Antenna System}
\newacronym{dash}{DASH}{Dynamic Adaptive Streaming over HTTP}
\newacronym{dc}{DC}{Dual Connectivity}
\newacronym{dccp}{DCCP}{Datagram Congestion Control Protocol}
\newacronym{dce}{DCE}{Direct Code Execution}
\newacronym{dci}{DCI}{Downlink Control Information}
\newacronym{dctcp}{DCTCP}{Data Center TCP}
\newacronym{dl}{DL}{Downlink}
\newacronym{dmr}{DMR}{Deadline Miss Ratio}
\newacronym{dmrs}{DMRS}{DeModulation Reference Signal}
\newacronym{drlcc}{DRL-CC}{Deep Reinforcement Learning Congestion Control}
\newacronym{drs}{DRS}{Discovery Reference Signal}
\newacronym{du}{DU}{Distributed Unit}
\newacronym{e2e}{E2E}{end-to-end}
\newacronym{ecaas}{ECaaS}{Edge-Cloud-as-a-Service}
\newacronym{ecn}{ECN}{Explicit Congestion Notification}
\newacronym{edf}{EDF}{Earliest Deadline First}
\newacronym{embb}{eMBB}{Enhanced Mobile Broadband}
\newacronym{empower}{EMPOWER}{EMpowering transatlantic PlatfOrms for advanced WirEless Research}
\newacronym{enb}{eNB}{evolved Node Base}
\newacronym{endc}{EN-DC}{E-UTRAN-\gls{nr} \gls{dc}}
\newacronym{epc}{EPC}{Evolved Packet Core}
\newacronym{eps}{EPS}{Evolved Packet System}
\newacronym{es}{ES}{Edge Server}
\newacronym{etsi}{ETSI}{European Telecommunications Standards Institute}
\newacronym[firstplural=Estimated Times of Arrival (ETAs)]{eta}{ETA}{Estimated Time of Arrival}
\newacronym{eutran}{E-UTRAN}{Evolved Universal Terrestrial Access Network}
\newacronym{faas}{FaaS}{Function-as-a-Service}
\newacronym{fapi}{FAPI}{Functional Application Platform Interface}
\newacronym{fdd}{FDD}{Frequency Division Duplexing}
\newacronym{fdm}{FDM}{Frequency Division Multiplexing}
\newacronym{fdma}{FDMA}{Frequency Division Multiple Access}
\newacronym{fed4fire}{FED4FIRE+}{Federation 4 Future Internet Research and Experimentation Plus}
\newacronym{fir}{FIR}{Finite Impulse Response}
\newacronym{fit}{FIT}{Future \acrlong{iot}}
\newacronym{fpga}{FPGA}{Field Programmable Gate Array}
\newacronym{fr2}{FR2}{Frequency Range 2}
\newacronym{fs}{FS}{Fast Switching}
\newacronym{fscc}{FSCC}{Flow Sharing Congestion Control}
\newacronym{ftp}{FTP}{File Transfer Protocol}
\newacronym{fw}{FW}{Flow Window}
\newacronym{ge}{GE}{Gaussian Elimination}
\newacronym{gnb}{gNB}{5G base station}
\newacronym{gop}{GOP}{Group of Pictures}
\newacronym{gpr}{GPR}{Gaussian Process Regressor}
\newacronym{gpu}{GPU}{Graphics Processing Unit}
\newacronym{gtp}{GTP}{GPRS Tunneling Protocol}
\newacronym{gtpc}{GTP-C}{GPRS Tunnelling Protocol Control Plane}
\newacronym{gtpu}{GTP-U}{GPRS Tunnelling Protocol User Plane}
\newacronym{gtpv2c}{GTPv2-C}{\gls{gtp} v2 - Control}
\newacronym{gw}{GW}{Gateway}
\newacronym{harq}{HARQ}{Hybrid Automatic Repeat reQuest}
\newacronym{hetnet}{HetNet}{Heterogeneous Network}
\newacronym{hh}{HH}{Hard Handover}
\newacronym{hol}{HOL}{Head-of-Line}
\newacronym{hqf}{HQF}{Highest-quality-first}
\newacronym{hss}{HSS}{Home Subscription Server}
\newacronym{http}{HTTP}{HyperText Transfer Protocol}
\newacronym{ia}{IA}{Initial Access}
\newacronym{iab}{IAB}{Integrated Access and Backhaul}
\newacronym{ic}{IC}{Incident Command}
\newacronym{ietf}{IETF}{Internet Engineering Task Force}
\newacronym{imsi}{IMSI}{International Mobile Subscriber Identity}
\newacronym{imt}{IMT}{International Mobile Telecommunication}
\newacronym{iot}{IoT}{Internet of Things}
\newacronym{ip}{IP}{Internet Protocol}
\newacronym{itu}{ITU}{International Telecommunication Union}
\newacronym{kpi}{KPI}{Key Performance Indicator}
\newacronym{kpm}{KPM}{Key Performance Measurement}
\newacronym{kvm}{KVM}{Kernel-based Virtual Machine}
\newacronym{los}{LOS}{Line-of-Sight}
\newacronym{lsm}{LSM}{Link-to-System Mapping}
\newacronym{lstm}{LSTM}{Long Short Term Memory}
\newacronym{lte}{LTE}{Long Term Evolution}
\newacronym{lxc}{LXC}{Linux Container}
\newacronym{m2m}{M2M}{Machine to Machine}
\newacronym{mac}{MAC}{Medium Access Control}
\newacronym{manet}{MANET}{Mobile Ad Hoc Network}
\newacronym{mano}{MANO}{Management and Orchestration}
\newacronym{mc}{MC}{Multi-Connectivity}
\newacronym{mcc}{MCC}{Mobile Cloud Computing}
\newacronym{mchem}{MCHEM}{Massive Channel Emulator}
\newacronym{mcs}{MCS}{Modulation and Coding Scheme}
\newacronym{mec}{MEC}{Multi-access Edge Computing}
\newacronym{mec2}{MEC}{Mobile Edge Cloud}
\newacronym{mfc}{MFC}{Mobile Fog Computing}
\newacronym{mgen}{MGEN}{Multi-Generator}
\newacronym{mi}{MI}{Mutual Information}
\newacronym{mib}{MIB}{Master Information Block}
\newacronym{miesm}{MIESM}{Mutual Information Based Effective SINR}
\newacronym{mimo}{MIMO}{Multiple Input, Multiple Output}
\newacronym{ml}{ML}{Machine Learning}
\newacronym{mlr}{MLR}{Maximum-local-rate}
\newacronym[plural=\gls{mme}s,firstplural=Mobility Management Entities (MMEs)]{mme}{MME}{Mobility Management Entity}
\newacronym{mmtc}{mMTC}{Massive Machine-Type Communications}
\newacronym{mmwave}{mmWave}{millimeter wave}
\newacronym{mpdccp}{MP-DCCP}{Multipath Datagram Congestion Control Protocol}
\newacronym{mptcp}{MPTCP}{Multipath TCP}
\newacronym{mr}{MR}{Maximum Rate}
\newacronym{mrdc}{MR-DC}{Multi \gls{rat} \gls{dc}}
\newacronym{mse}{MSE}{Mean Square Error}
\newacronym{mss}{MSS}{Maximum Segment Size}
\newacronym{mt}{MT}{Mobile Termination}
\newacronym{mtd}{MTD}{Machine-Type Device}
\newacronym{mtu}{MTU}{Maximum Transmission Unit}
\newacronym{mumimo}{MU-MIMO}{Multi-user \gls{mimo}}
\newacronym{mvno}{MVNO}{Mobile Virtual Network Operator}
\newacronym{nalu}{NALU}{Network Abstraction Layer Unit}
\newacronym{nas}{NAS}{Non-Access Stratum}
\newacronym{nbiot}{NB-IoT}{Narrow Band IoT}
\newacronym{nfv}{NFV}{Network Function Virtualization}
\newacronym{nfvi}{NFVI}{Network Function Virtualization Infrastructure}
\newacronym{ngrg}{nGRG}{next Generation Research Group}
\newacronym{ni}{NI}{Network Interfaces}
\newacronym{nic}{NIC}{Network Interface Card}
\newacronym{nlos}{NLOS}{Non-Line-of-Sight}
\newacronym{now}{NOW}{Non Overlapping Window}
\newacronym{nsm}{NSM}{Network Service Mesh}
\newacronym{nr}{NR}{New Radio}
\newacronym{nrf}{NRF}{Network Repository Function}
\newacronym{nsa}{NSA}{Non Stand Alone}
\newacronym{nse}{NSE}{Network Slicing Engine}
\newacronym{nssf}{NSSF}{Network Slice Selection Function}
\newacronym{o2i}{O2I}{Outdoor to Indoor}
\newacronym{oai}{OAI}{OpenAirInterface}
\newacronym{oaicn}{OAI-CN}{\gls{oai} \acrlong{cn}}
\newacronym{oairan}{OAI-RAN}{\acrlong{oai} \acrlong{ran}}
\newacronym{oam}{OAM}{Operations, Administration and Maintenance}
\newacronym{ofdm}{OFDM}{Orthogonal Frequency Division Multiplexing}
\newacronym{olia}{OLIA}{Opportunistic Linked Increase Algorithm}
\newacronym{omec}{OMEC}{Open Mobile Evolved Core}
\newacronym{onap}{ONAP}{Open Network Automation Platform}
\newacronym{onf}{ONF}{Open Networking Foundation}
\newacronym{onos}{ONOS}{Open Networking Operating System}
\newacronym{oom}{OOM}{\gls{onap} Operations Manager}
\newacronym{opnfv}{OPNFV}{Open Platform for \gls{nfv}}
\newacronym{oran}{O-RAN}{Open Radio Access Network}
\newacronym{orbit}{ORBIT}{Open-Access Research Testbed for Next-Generation Wireless Networks}
\newacronym{os}{OS}{Operating System}
\newacronym{oss}{OSS}{Operations Support System}
\newacronym{otic}{OTIC}{Open Testing \& Integration Centre}
\newacronym{pa}{PA}{Position-aware}
\newacronym{pase}{PASE}{Prioritization, Arbitration, and Self-adjusting Endpoints}
\newacronym{pawr}{PAWR}{Platforms for Advanced Wireless Research}
\newacronym{pbch}{PBCH}{Physical Broadcast Channel}
\newacronym{pcef}{PCEF}{Policy and Charging Enforcement Function}
\newacronym{pcfich}{PCFICH}{Physical Control Format Indicator Channel}
\newacronym{pcrf}{PCRF}{Policy and Charging Rules Function}
\newacronym{pdcch}{PDCCH}{Physical Downlink Control Channel}
\newacronym{pdcp}{PDCP}{Packet Data Convergence Protocol}
\newacronym{pdsch}{PDSCH}{Physical Downlink Shared Channel}
\newacronym{pdu}{PDU}{Packet Data Unit}
\newacronym{pf}{PF}{Proportional Fair}
\newacronym{pgw}{PGW}{Packet Gateway}
\newacronym{phich}{PHICH}{Physical Hybrid ARQ Indicator Channel}
\newacronym{phy}{PHY}{Physical}
\newacronym{pmch}{PMCH}{Physical Multicast Channel}
\newacronym{pmi}{PMI}{Precoding Matrix Indicators}
\newacronym{powder}{POWDER}{Platform for Open Wireless Data-driven Experimental Research}
\newacronym{ppo}{PPO}{Proximal Policy Optimization}
\newacronym{ppp}{PPP}{Poisson Point Process}
\newacronym{prach}{PRACH}{Physical Random Access Channel}
\newacronym{prb}{PRB}{Physical Resource Block}
\newacronym{psnr}{PSNR}{Peak Signal to Noise Ratio}
\newacronym{pss}{PSS}{Primary Synchronization Signal}
\newacronym{pucch}{PUCCH}{Physical Uplink Control Channel}
\newacronym{pusch}{PUSCH}{Physical Uplink Shared Channel}
\newacronym{rar}{RAR}{Random Access Response}
\newacronym{qam}{QAM}{Quadrature Amplitude Modulation}
\newacronym{qci}{QCI}{\gls{qos} Class Identifier}
\newacronym{5qi}{5QI}{5G \gls{qos} Identifier}
\newacronym{qoe}{QoE}{Quality of Experience}
\newacronym{qos}{QoS}{Quality of Service}
\newacronym{quic}{QUIC}{Quick UDP Internet Connections}
\newacronym{rach}{RACH}{Random Access Channel}
\newacronym{ran}{RAN}{Radio Access Network}
\newacronym[firstplural=Radio Access Technologies (RATs)]{rat}{RAT}{Radio Access Technology}
\newacronym{rcn}{RCN}{Research Coordination Network}
\newacronym{rc}{RC}{RAN Control}
\newacronym{rec}{REC}{Radio Edge Cloud}
\newacronym{red}{RED}{Random Early Detection}
\newacronym{renew}{RENEW}{Reconfigurable Eco-system for Next-generation End-to-end Wireless}
\newacronym{rf}{RF}{Radio Frequency}
\newacronym{rfc}{RFC}{Request for Comments}
\newacronym{rfr}{RFR}{Random Forest Regressor}
\newacronym{ric}{RIC}{RAN Intelligent Controller}
\newacronym{rlc}{RLC}{Radio Link Control}
\newacronym{rlf}{RLF}{Radio Link Failure}
\newacronym{rlnc}{RLNC}{Random Linear Network Coding}
\newacronym{rmr}{RMR}{RIC Message Router}
\newacronym{rmse}{RMSE}{Root Mean Squared Error}
\newacronym{rnis}{RNIS}{Radio Network Information Service}
\newacronym{rr}{RR}{Round Robin}
\newacronym{rrc}{RRC}{Radio Resource Control}
\newacronym{rrm}{RRM}{Radio Resource Management}
\newacronym{rru}{RRU}{Remote Radio Unit}
\newacronym{rsrp}{RSRP}{Reference Signal Received Power}
\newacronym{rsrq}{RSRQ}{Reference Signal Received Quality}
\newacronym{rss}{RSS}{Received Signal Strength}
\newacronym{rssi}{RSSI}{Received Signal Strength Indicator}
\newacronym{rtt}{RTT}{Round Trip Time}
\newacronym{ru}{RU}{Radio Unit}
\newacronym{rw}{RW}{Receive Window}
\newacronym{rx}{RX}{Receiver}
\newacronym{s1ap}{S1AP}{S1 Application Protocol}
\newacronym{sa}{SA}{standalone}
\newacronym{sack}{SACK}{Selective Acknowledgment}
\newacronym{sap}{SAP}{Service Access Point}
\newacronym{sc2}{SC2}{Spectrum Collaboration Challenge}
\newacronym{scef}{SCEF}{Service Capability Exposure Function}
\newacronym{sch}{SCH}{Secondary Cell Handover}
\newacronym{scoot}{SCOOT}{Split Cycle Offset Optimization Technique}
\newacronym{sctp}{SCTP}{Stream Control Transmission Protocol}
\newacronym{sdap}{SDAP}{Service Data Adaptation Protocol}
\newacronym{sdk}{SDK}{Software Development Kit}
\newacronym{sdm}{SDM}{Space Division Multiplexing}
\newacronym{sdma}{SDMA}{Spatial Division Multiple Access}
\newacronym{sdn}{SDN}{Software-defined Networking}
\newacronym{sdr}{SDR}{Software-defined Radio}
\newacronym{seba}{SEBA}{SDN-Enabled Broadband Access}
\newacronym{sgsn}{SGSN}{Serving GPRS Support Node}
\newacronym{sgw}{SGW}{Service Gateway}
\newacronym{si}{SI}{Study Item}
\newacronym{sib}{SIB}{Secondary Information Block}
\newacronym{sinr}{SINR}{Signal to Interference plus Noise Ratio}
\newacronym{sip}{SIP}{Session Initiation Protocol}
\newacronym{siso}{SISO}{Single Input, Single Output}
\newacronym{sla}{SLA}{Service Level Agreement}
\newacronym{sm}{SM}{Service Model}
\newacronym{smf}{SMF}{Session Management Function}
\newacronym{smo}{SMO}{Service Management and Orchestration}
\newacronym{sms}{SMS}{Short Message Service}
\newacronym{smsgmsc}{SMS-GMSC}{\gls{sms}-Gateway}
\newacronym{snr}{SNR}{Signal-to-Noise-Ratio}
\newacronym{cnr}{CNR}{Carrier-to-Noise-Ratio}
\newacronym{son}{SON}{Self-Organizing Network}
\newacronym{sptcp}{SPTCP}{Single Path TCP}
\newacronym{srb}{SRB}{Service Radio Bearer}
\newacronym{srn}{SRN}{Standard Radio Node}
\newacronym{srs}{SRS}{Sounding Reference Signal}
\newacronym{prs}{SRS}{Positioning Reference Signal}
\newacronym{zc}{ZC}{Zadoff-Chu}
\newacronym{ta}{TA}{Timing Advance}
\newacronym{tac}{TA-Command}{Timing Advance Command}
\newacronym{ss}{SS}{Synchronization Signal}
\newacronym{ssb}{SSB}{Synchronization Signal Block}
\newacronym{sss}{SSS}{Secondary Synchronization Signal}
\newacronym{st}{ST}{Spanning Tree}
\newacronym{svc}{SVC}{Scalable Video Coding}
\newacronym{tb}{TB}{Transport Block}
\newacronym{tcp}{TCP}{Transmission Control Protocol}
\newacronym{tdd}{TDD}{Time Division Duplexing}
\newacronym{tdm}{TDM}{Time Division Multiplexing}
\newacronym{tdma}{TDMA}{Time Division Multiple Access}
\newacronym{tfl}{TfL}{Transport for London}
\newacronym{tfrc}{TFRC}{TCP-Friendly Rate Control}
\newacronym{tft}{TFT}{Traffic Flow Template}
\newacronym{tgen}{TGEN}{Traffic Generator}
\newacronym{tip}{TIP}{Telecom Infra Project}
\newacronym{tm}{TM}{Transparent Mode}
\newacronym{to}{TO}{Telco Operator}
\newacronym{tr}{TR}{Technical Report}
\newacronym{trp}{TRP}{Transmitter Receiver Pair}
\newacronym{ts}{TS}{Technical Specification}
\newacronym{tti}{TTI}{Transmission Time Interval}
\newacronym{ttt}{TTT}{Time-to-Trigger}
\newacronym{tx}{TX}{Transmitter}
\newacronym{uas}{UAS}{Unmanned Aerial System}
\newacronym{uav}{UAV}{Unmanned Aerial Vehicle}
\newacronym{udm}{UDM}{Unified Data Management}
\newacronym{udp}{UDP}{User Datagram Protocol}
\newacronym{udr}{UDR}{Unified Data Repository}
\newacronym{ue}{UE}{User Equipment}
\newacronym{uhd}{UHD}{\gls{usrp} Hardware Driver}
\newacronym{ul}{UL}{Uplink}
\newacronym{jcas}{JCAS}{Joint Communication and Sensing}
\newacronym{isac}{ISAC}{Integrated Sensing and Communication}
\newacronym{um}{UM}{Unacknowledged Mode}
\newacronym{uml}{UML}{Unified Modeling Language}
\newacronym{upa}{UPA}{Uniform Planar Array}
\newacronym{upf}{UPF}{User Plane Function}
\newacronym{urllc}{URLLC}{Ultra Reliable and Low Latency Communications}
\newacronym{usa}{U.S.}{United States}
\newacronym{usim}{USIM}{Universal Subscriber Identity Module}
\newacronym{usrp}{USRP}{Universal Software Radio Peripheral}
\newacronym{utc}{UTC}{Urban Traffic Control}
\newacronym{vim}{VIM}{Virtualization Infrastructure Manager}
\newacronym{vm}{VM}{Virtual Machine}
\newacronym{vnf}{VNF}{Virtual Network Function}
\newacronym{volte}{VoLTE}{Voice over \gls{lte}}
\newacronym{voltha}{VOLTHA}{Virtual OLT HArdware Abstraction}
\newacronym{vr}{VR}{Virtual Reality}
\newacronym{vran}{vRAN}{Virtualized \gls{ran}}
\newacronym{vss}{VSS}{Video Streaming Server}
\newacronym{wbf}{WBF}{Wired Bias Function}
\newacronym{wf}{WF}{Waterfilling}
\newacronym{wg}{WG}{Working Group}
\newacronym{wlan}{WLAN}{Wireless Local Area Network}
\newacronym{osm}{OSM}{Open Source Management and Orchestration}
\newacronym{pnf}{PNF}{Physical Network Function}
\newacronym{drl}{DRL}{Deep Reinforcement Learning}
\newacronym{mtc}{MTC}{Machine-type Communications}
\newacronym{osc}{OSC}{O-RAN Software Community}
\newacronym{mns}{MnS}{Management Services}
\newacronym{ves}{VES}{\gls{vnf} Event Stream}
\newacronym{ei}{EI}{Enrichment Information}
\newacronym{fh}{FH}{Fronthaul}
\newacronym{fft}{FFT}{Fast Fourier Transform}
\newacronym{laa}{LAA}{Licensed-Assisted Access}
\newacronym{plfs}{PLFS}{Physical Layer Frequency Signals}
\newacronym{ptp}{PTP}{Precision Time Protocol}
\newacronym{asic}{ASIC}{Application-specific Integrated Circuit}
\newacronym{aal}{AAL}{Acceleration Abstraction Layer}
\newacronym{fec}{FEC}{Forward Error Correction}
\newacronym{sdl}{SDL}{Shared Data Layer}
\newacronym{nib}{NIB}{Network Information Base}
\newacronym{rnib}{R-NIB}{RAN \gls{nib}}
\newacronym{fcaps}{FCAPS}{Fault, Configuration, Accounting, Performance, Security}
\newacronym{ie}{IE}{Information Element}
\newacronym{fg}{FG}{Focus Group}
\newacronym{osfg}{OSFG}{Open Source Focus Group}
\newacronym{sdfg}{SDFG}{Standard Development Focus Group}
\newacronym{tifg}{TIFG}{Test \& Integration Focus Group}
\newacronym{sfg}{SFG}{Security Focus Group}
\newacronym{swg}{SWG}{Security Work Group}
\newacronym{e2sm}{E2SM}{E2 Service Model}
\newacronym{e2ap}{E2AP}{E2 Application Potocol}
\newacronym{tsc}{TSC}{Technical Steering Committee}
\newacronym{sdo}{SDO}{Standard-Development Organization}
\newacronym{sql}{SQL}{Structured Query Language}
\newacronym{ssh}{SSH}{Secure Shell}
\newacronym{tls}{TLS}{Transport Layer Security}
\newacronym{netconf}{NETCONF}{Network Configuration Protocol}
\newacronym{dtls}{DTLS}{Datagram Transport Layer Security}
\newacronym{cmp}{CMP}{Certificate Management Protocol}
\newacronym{ccc}{CCC}{Cell Configuration and Control}
\newacronym{dsp}{DSP}{Digital Signal Processing}
\newacronym{opex}{OPEX}{Operational Expenses}
\newacronym{cbrs}{CBRS}{Citizen Broadband Radio Service}
\newacronym{ntn}{NTN}{Non-terrestrial Network}
\newacronym{gbr}{GBR}{Guaranteed Bitrate}
\newacronym{sps}{SPS}{Semi-Persistent Scheduling}
\newacronym{tbs}{TBS}{Transport Block Size}
\newacronym{gnss}{GNSS}{Global Navigation Satellite System}
\newacronym{tof}{ToF}{Time of Flight}
\newacronym{rtof}{RToF}{Return Time of Flight}
\newacronym{rsig}{RS}{Reference Signal}
\newacronym{nrtric}{near-RT RIC}{near-Real Time Ran Intelligent Controller}
\newacronym{nonrtric}{non-RT RIC}{non-Real Time Ran Intelligent Controller}
\newacronym{aoa}{AoA}{Angle of Arrival}
\newacronym{tdoa}{TDoA}{Time Difference of Arrival}
\newacronym{rtoa}{RToA}{Return Time of Arrival}
\newacronym{ecdf}{ECDF}{Empirical Cumulative Distribution Function}
\newacronym{ris}{RIS}{Reconfigurable Intelligent Surface}
\newacronym{srd}{SRD}{Smart Radio Device}
\newacronym{gfbr}{GFBR}{Guaranteed Flow Bit Rate}
\newacronym{rg}{RG}{Resource Grid}
\newacronym{rb}{RB}{Resource Block}
\newacronym{re}{RE}{Resource Element}
\newacronym{rfra}{RF}{Radio Frame}
\newacronym{scs}{SCS}{Subcarrier Spacing}
\newacronym{ec}{EC}{Edge Computing}
\newacronym{5g-phy}{5G PHY}{5G Physical Layer}
\newacronym{fr1}{FR1}{Frequency Range 1}
\newacronym{tpm}{tpm}{times per minute}
\newacronym{ra}{RA}{Random Access}
\newacronym{crnti}{C-RNTI}{Cell Radio Network Temporary Identifier}
\newacronym{rnti}{RNTI}{Radio Network Temporary Identifier}
\newacronym{toa}{ToA}{Time of Arrival}
\newacronym{kf}{KF}{Kalman Filter}
\newacronym{ekf}{EKF}{Extended Kalman Filter}
\newacronym{nls}{NLS}{Nonlinear Least Squares}
\newacronym{lmf}{LMF}{Location Management Function}
\newacronym{gdop}{GDOP}{Geometric Dilution of Precision}
\newacronym{pbuf}{protobuf}{Protocol Buffers}

%% file: Content/introTEST.tex
\section{Introduction}
\label{sc:intro}

Accurate and reliable positioning is becoming an important requirement for mobile networks. Emerging applications, including emergency response, industrial automation, extended reality, asset tracking, and network-aware services, require location estimates in indoor and dense urban environments. However, \glspl{gnss}, which remain the main source of global positioning information, provide limited performance under poor satellite visibility.

This limitation has motivated the integration of sensing and positioning capabilities into cellular systems. The broader vision of \gls{isac} is to reuse communication infrastructure, spectrum, and waveforms to provide sensing and positioning services beyond data transmission~\cite{gonzalez_prelcic2024isac,etsi_isc001}. Within 5G, positioning has been standardized through the \gls{lmf} and dedicated \gls{rsig} procedures. These mechanisms represent an important step toward network-integrated localization, but their practical deployment remains constrained. In particular, downlink positioning requires the \gls{ue} to perform measurements and report them to the network, making the service dependent on chipset support, device implementation choices, and additional \gls{ue}-side processing. For operators, this is a critical limitation: a localization service cannot be deployed entirely from the network side if its availability depends on whether each commercial device exposes the required measurement capabilities~\cite{cotsLocalizationFeasibility}.

Uplink positioning offers a complementary opportunity. Commercial \glspl{ue} already transmit \glspl{srs} for channel sounding, and these signals have correlation properties that make them suitable for timing-based ranging. If multiple geographically distributed \glspl{gnb} could observe the same uplink transmission, the network could estimate the \gls{ue} position without requiring additional localization functionality at the device. This would make positioning an infrastructure-side service: transparent to the \gls{ue}, controlled by the operator, and configurable according to the accuracy, latency, and resource requirements of different applications. From a broader perspective, network-native location information could feed the \gls{ai} layer of future \gls{oran}-based architectures, where \gls{ai}-driven functions would exploit it to optimize \gls{ran} operations. At the same time, controlled application interfaces could expose positioning services to external applications, enabling use cases such as industrial automation and context-aware edge services.

The main limitation is that conventional \glspl{ran} do not expose the information required for cooperative uplink processing. A serving \gls{gnb} knows the scheduling decisions, resource allocations, and reference sequences associated with its \glspl{ue}. A neighboring \gls{gnb} may receive the same \gls{srs}, but it does not know when it is scheduled or which sequence must be used for coherent processing. Therefore, multiple receptions of the same uplink signal cannot directly be transformed into a set of cooperative ranging measurements.

The \gls{oran} architecture provides the programmability required to address this problem. The \gls{oran} Alliance promotes disaggregating \glspl{ran} components and using open, standardized interfaces~\cite{oran_polese}. In particular, the near-RT RIC enables software-based control and inference through network microservices implemented as xApps. A growing body of research has leveraged this openness to extend \gls*{ran} functionalities beyond conventional signal processing. Previous works have used this architecture to detect and mitigate interference~\cite{reus2023senseoran} and physical-layer attacks~\cite{wen20245g}, to use spectrum unconventionally~\cite{lizarribar2024oran}, as well as to support sensing infrastructures~\cite{pan2025openran_sensing}. These studies show that \gls{oran} can extend \gls{ran} functionality beyond conventional communication operations, allowing the network to dynamically adapt its behavior.

Based on these capabilities, we introduce RUN-O-RAN, an \gls{oran}-based 5G positioning micro-service that estimates the location of commercial \glspl{ue} from standard uplink \gls{srs} transmissions. Unlike previous \gls{oran} positioning applications~\cite{zeng2022new,ko2021beamforming,lizarribar2024oransense,mimoric2024}, RUN-O-RAN requires no additional hardware, positioning procedures, or computation at the \gls{ue}, as the network performs the full localization process while the user equipment remains fully standards-compliant.

Recent work has demonstrated that \gls{3gpp}-compliant \gls{ul}-\gls{tdoa} positioning can be integrated into practical open-source 5G implementations~\cite{malik2025oai_ultdoa}. These solutions establish the feasibility of uplink timing-based localization, but they rely on the conventional \gls{lmf}-based architecture, limiting the applicability to intra-\gls{du} deployments, in which multiple \glspl{ru} are controlled by the same \gls{du} (\gls{oran} functional split 7-2x). Extending this approach across independent \glspl{gnb} remains an open challenge because the standard \gls{ran}-\gls{cn} architecture provides no general mechanism for distributing the scheduling information and reference signal configuration required for neighboring \glspl{gnb} to process \gls{ul} transmissions from non-associated \glspl{ue}. 

RUN-O-RAN addresses this limitation by introducing an xApp running in the \gls{nrtric} that orchestrates cooperative uplink localization across multiple gNBs. To the best of our knowledge, this is the first fully network-based uplink positioning system integrated into the \gls{ran} and capable of coordinating independent \glspl{gnb} to localize \gls{cots} \glspl{ue}, establishing network-controlled positioning as a native \gls{ran} capability and enabling a broad range of location-aware applications. Through the E2 interface, the xApp collects and distributes the information required to coordinate the serving and neighboring \glspl{gnb} for positioning, including anchor identities and coordinates, negotiated \gls{srs} configurations, reference sequences, timing-advance updates, and received signal observations. This enables neighboring \glspl{gnb} to process \gls{srs} transmissions from non-associated \glspl{ue} and operate as localization anchors without requiring modifications to either the \gls{ue} or the underlying \gls{3gpp} signaling procedures.

Realizing cooperative uplink localization requires addressing three main challenges. First, the network must provide non-serving \glspl{gnb} with the information required to process uplink \glspl{srs}. Second, the measured \gls{srs} delays must be corrected before they can be interpreted as geometric ranges, since synchronization offsets, clock drift, timing-advance updates, and multipath affect the estimated propagation time. Finally, the complete system must be validated under realistic yet controlled propagation conditions while preserving standards-compliant \gls{oran} operation. To address these challenges, this paper makes the following novel contributions:
 \begin{itemize}
     \item We design and implement RUN-O-RAN, an \gls{oran}-native localization framework that enables serving and neighboring \glspl{gnb} to cooperate through a near-RT RIC xApp. The framework provides network-side uplink trilateration without modifications to commercial \glspl{ue}.
     \item We develop an \gls{srs}-based localization pipeline that extracts \gls{toa} measurements through cross-correlation with the known \gls{srs} sequence, compensates for timing-advance and clock-drift components, and estimates the \gls{ue} position using geometric and Kalman-based techniques.
     \item We implement and evaluate RUN-O-RAN using a hybrid experimental testbed composed of a COTS 5G \gls{ue}, OpenAirInterface, Open5GS, a USRP N310, a Keysight PROPSIM channel emulator, and \gls{oran}-compatible \gls{gnb} instances. The evaluation includes $288{,}000$ \gls{srs} transmissions collected in static and dynamic urban scenarios.
\end{itemize}
The remainder of the paper is organized as follows. Section~\ref{sc: background} introduces the physical-layer foundations of uplink positioning. Section~\ref{sc: run-o-ran} presents the RUN-O-RAN architecture and localization pipeline, while Section~\ref{sc: setup} describes the experimental platform. Section~\ref{sc: results} reports the evaluation results, and Section~\ref{sc: discussion} discusses limitations and future research directions. Finally, Section~\ref{sc: conclusion} concludes the paper.


%% file: Content/background.tex
\input{Content/figures/flowcharts/PHYgrid}

\section{5G NR Background for Uplink Positioning} 
\label{sc: background}
In this section, we provide an overview of 5G radio resources' organization, synchronization procedures, and \gls*{srs} configuration to understand the design choices we made to enable a reliable \gls*{ul} \gls*{ue} positioning.
In this work, we focus on \gls*{nr} operating in the \gls*{fr1}. Although some parameter values differ for other frequency ranges, the overall scheme remains unchanged.

\subsection{Resource Configuration}%
\label{sbs: back-resource configuration}
\gls*{5g} has been designed with the unprecedented goal of unifying diverse verticals under a single, reconfigurable access network. 
To accommodate the broad spectrum of \gls{5g} applications, the radio interface introduces a highly flexible time–frequency structure defined by a set of configurations known as numerologies: $\mu = \{0, 1,  ...,6\}$, which allow the network to adapt the physical layer to diverse latency and coverage needs. Low numerologies (e.g., $\mu=0$) with narrow subcarrier spacing are typically used for wide-area, delay-tolerant services, while higher numerologies (e.g., $\mu \geq 3$) favor low-latency, high-throughput applications such as vehicular or industrial communications.

In the time domain, each numerology corresponds to a specific radio frame configuration as defined by \gls*{3gpp} specifications \cite{ts38211}. The radio frames have a fixed length of $10$~ms, divided into $10$ sub-frames of $1$~ms each. The number of slots per sub-frame scales exponentially with the numerology, following: $N_{\textit{slot}}^{\textit{subframe}}(\mu) = 2^\mu = \{1, 2, 4, 8, 16, 32, 64\}$. Moreover, each slot comprises  $N_{symb}=14$ \gls*{ofdm} symbols with a normal \gls*{cyp} or $N_{symb}=12$ symbols with an extended \gls*{cyp}. 

In the frequency domain, the numerology determines the \gls*{scs}, defined as $SCS = 2^\mu \cdot 15 \text{~kHz}$. 
The maximum achievable bandwidth $B$ that can be obtained for a given number of sub-carriers $N_{\textit{SC}}$ is computed as $B = N_{\textit{SC}} \cdot SCS = N_{\textit{SC}} \cdot (2^\mu \cdot 15 \text{~kHz})$. This parameterization allows the network to finely balance spectral efficiency and coverage, adapting to the propagation conditions and service constraints of each deployment scenario.

The \gls*{rg} is modeled and managed in both the time and frequency domains, subdivided into \glspl*{rb}, each comprising one \gls*{ofdm} symbol in the time domain and $N_{\textit{sc}}^{\textit{RB}} = 12$ sub-carriers in the frequency domain, regardless of the \gls*{cyp} configuration. Thus, the \gls*{prb} duration is $T^{\textit{RB}} = 2^{-\mu} $~ms and its bandwidth is $B^{\textit{RB}} = N_{\textit{sc}}^{\textit{RB}} \cdot SCS = 12 \cdot 2^\mu \cdot 15$~kHz. The smallest allocable resource unit is the \gls*{re}, corresponding to one \gls*{ofdm} symbol in the time domain and one sub-carrier in the frequency domain. 
Fig.~\ref{fig:resource_grid} illustrates a \gls*{rg}, with the x-axis representing the time domain and the y-axis the frequency domain. As depicted, given a total bandwidth $B$, the total number of \glspl*{rb} is $N_{RB} = \nicefrac{B}{\left(N_{\textit{sc}}^{\textit{RB}} \cdot SCS\right)}$.

\input{Content/figures/flowcharts/timing_advance}

\subsection{\gls{gnb}-\gls{ue} Synchronization}%
\label{sbs: back-synchronization}
When a \gls{ue} attaches to a \gls{gnb}, it aligns its internal time reference to the received \gls{dl} \gls{ssb}. Due to the propagation delay $\tau_{\text{\tiny{UE,gNB}}}$, a timing offset exists between the \gls{ue} and \gls{gnb} reference times. As a result, in the absence of compensation, \gls{ul} transmissions would be received at the \gls{gnb} with a total delay of $2\cdot\tau_{\text{\tiny{UE,gNB}}}$, accounting for both \gls{dl} and \gls{ul} propagation. For sufficiently large distances, this delay may cause \gls{ul} signals to fall outside the intended time slot boundaries, leading to inter-symbol interference. To prevent this effect, \gls{3gpp} specifies a closed-loop \gls{ta} mechanism~\cite{ts38211, ts38213}, whereby the serving \gls{gnb} periodically estimates the timing offset and commands the \gls{ue} to adjust its transmission timing accordingly.

In particular, the \gls*{ul} reception delay is periodically estimated and quantized in steps of
$1024 \cdot 2^{-\mu}\cdot T_c$, 
where $T_c=0.509$~ns denotes the basic time unit~\cite{ts38211}.
At each $k$-th update cycle, a \gls{tac} with value $\mathrm{TA}_k \in \{ 0,1,...,63\}$ is reported to the \gls{ue}.
The \gls{ue} is required to apply the corresponding timing adjustment within $6$ slots by advancing or delaying its subsequent transmissions according to
\begin{equation}
    \Delta \tau_{\text{\tiny{TA}}}^k =
    (\mathrm{TA}_k - 31)\cdot 1024 \cdot 2^{-\mu} \cdot T_c,
\end{equation}
where $\Delta \tau_{\text{\tiny{TA}}}^k$ denotes the incremental timing adjustment applied at the $k$-th update, which depends on the received $\mathrm{TA}_k$ value and the numerology index $\mu$~\cite{ts38321}.
Since each \gls{tac} is computed relative to the previously adjusted time reference, the resulting timing control process is inherently cumulative.

This iterative procedure forms a closed-loop timing adjustment that keeps the \gls{ul} transmissions of each \gls{ue} aligned with the \gls{gnb}’s reference time.  More precisely, after the $k$-th update, the cumulative timing advance is given by:
\begin{equation} \label{eq: tau}
    \tau_{\text{\tiny{TA}}}^k = 
    \tau_{\text{\tiny{TA}}}^0+
    \sum_{j=1}^{k}{\Delta \tau_{\text{\tiny{TA}}}^j},
\end{equation}
where $\tau_{\text{\tiny{TA}}}^0$ is the \gls{rar}-\gls{ta}\footnote{\gls{rar}-\gls{ta} is computed as $(N^{\mathrm{TA},\text{off}}+\mathrm{TA}_0\cdot 1024 \cdot 2^{-\mu})\cdot T_c$ with $N^{\mathrm{TA},\text{off}}$ being a fixed cell configuration parameter that is used to adapt the random access \gls{ta} to the specific geometry of the cell~(\cite{ts38133,ts38211}).} and $k$ is the number of \glspl*{tac} received by the \gls*{ue} from its cell attachment up to that time. At the $k$-th adjustment the \gls{gnb} will therefore receive the \gls{ue}'s \gls{ul} transmissions with a delay of $2 \cdot \tau_{\text{\tiny{UE,gNB}}} + \tau_{\text{\tiny{TA}}}^k$.

In the context of \gls{ul}-based positioning, any loss or corruption of \glspl*{tac} invalidates the \gls{gnb} knowledge of the \gls{ue} time reference, making position estimation unfeasible. Therefore, continuous tracking of $\Delta \tau_{\text{\tiny{TA}}}^k$, 
along with the detection and compensation of missed \gls{ta} updates, is indispensable to the correct operation of the localization service.

\subsection{Reference Signals}%
\label{sbs: back-reference signals}
In 5G \gls*{nr}, \acrlong{rsig}s play a crucial role in ensuring efficient and reliable communication between the \gls*{ue} and the \gls*{gnb}. They are responsible for channel estimation, synchronization, measurements for mobility, and beam management \cite{ts38211}. The configuration on the \gls*{rg} of these reference signals is customizable, depending on the deployment scenario and service requirements, offering flexibility in network resource allocation.

Each \gls*{rsig} is designed for a specific purpose. For high-accuracy localization, the Positioning \gls*{rsig} (PRS) is the primary standardized option. PRS operates in the downlink and can achieve meter-level accuracy, on the order of a few meters with $100$~MHz bandwidth \cite{palama20245g}. However, it requires the \gls*{ue} to perform additional processing and to report positioning measurements or estimates back to the network, increasing both device complexity and signaling overhead. 

In this work, we instead target a seamless integration of localization within the network infrastructure. To this end, we avoid relying on PRS and focus on \gls{ul} \glspl*{rsig}, specifically the \gls*{srs}. This choice is motivated by the favorable auto-correlation properties of SRS signals: in particular, SRS sequences are derived from \gls{zc} sequences, which are complex-valued, constant-amplitude signals exhibiting ideal periodic autocorrelation, i.e., zero autocorrelation for all non-zero time shifts~\cite{hua2014analysis}.

\glspl*{srs} are primarily designed for \gls{ul} channel quality estimation, enabling the \gls*{gnb} to perform efficient scheduling and link adaptation. Their configuration is determined by the \gls*{gnb} according to the \gls*{ue}'s capabilities and is conveyed via \gls*{rrc} setup or reconfiguration messages. Specifically, the \textit{SRS-Config} and within that the \textit{SRS-Resource} parameters define the allocation of \glspl*{srs} within the \gls*{rg}.

In the frequency domain, the \gls*{srs} is transmitted starting from a given subcarrier $k_0$, extending for the upper $m_{SRS,b}$ \glspl*{prb}, occupying one every $K_{TC}$ subcarriers of those \glspl*{prb}. This results in a bandwidth occupation of
$B_{SRS} = m_{SRS,b} \cdot B^{RB}$~kHz 
and a total of $N_{SC}^{SRS} = m_{SRS,b} \cdot N_{SC}^{RB} / K_{TC}$ occupied subcarriers, each carrying an \gls*{ofdm} symbol. 
In the time domain, the signal's transmission starts at a given \gls*{ofdm} symbol $l_0 = N_{symb} -1 -l_{offset}$ of the slot and continues for \textit{nr\_of\_symbols} symbols.
The black-filled squares of Fig.~\ref{fig:resource_grid} depict an example of resource configuration for the \gls*{srs}. In this case, the parameters are: $K_{TC}=2$ as the signal is sent every other subcarrier, $l_{offset} = 4$ as the signal is sent on the fourth-to-last symbol, and $nr\_of\_symbols=1$ as it occupies only one \gls*{ofdm} symbol. 

%% file: Content/figures/flowcharts/PHYgrid.tex
\begin{figure}[t!]
    \centering

    \begin{tikzpicture}[scale = 0.6]
    
        \draw[step=0.5cm,black,very thin] (0,0) grid (7,8);
        \draw[step=0.5cm,black,very thin] (0,8.99) grid (7,10);
        \draw [very thin]  (0,8) -- (0,9);
        \draw [very thin]  (7,8) -- (7,9);
        \draw [dotted, very thick] (3.5,8.25) -- (3.5, 8.75);

        \fill[black] (5,0) rectangle (5.5,0.5);
        \fill[black] (5,1) rectangle (5.5,1.5);
        \fill[black] (5,2) rectangle (5.5,2.5);
        \fill[black] (5,3) rectangle (5.5,3.5);
        \fill[black] (5,4) rectangle (5.5,4.5);
        \fill[black] (5,5) rectangle (5.5,5.5);
        \fill[black] (5,6) rectangle (5.5,6.5);
        \fill[black] (5,7) rectangle (5.5,7.5);
        \fill[black] (5,9) rectangle (5.5,9.5);

        \draw [dashed, thin]  (-0.5,0.25) -- (0.25,0.25);
        \draw [dashed, thin]  (-0.5,0.75) -- (0.25,0.75);
        \draw [<->,thick] (-0.5,0.25) -- (-0.5,0.25 |- 0.25,0.75) node [above, rotate=90, midway] {\tiny $SCS$};

        \draw [very thick] (6.5,1) rectangle (7,7);
        \draw [very thick] (0,1) rectangle (0.5,7);
        \draw [very thin] (7,3.5) -- (7,4) node [anchor=west, midway] {\tiny Block};
        \draw [very thin] (7,4) -- (7,4.5) node [anchor=west, midway] {\tiny Resource};
        \draw [dashed, thin]  (-0.5,1) -- (0,1);
        \draw [dashed, thin]  (-0.5,7) -- (0,7);
        \draw [<->,thick] (-0.5,1) -- (-0.5,1 |- -0.5,7) node [above, rotate=90, midway] {\tiny $N_{\textit{sc}}^{\textit{RB}}$ subcarriers};


        \draw[dashed, thin] (-1.5,0) -- (0,0);
        \draw[dashed, thin] (-1.5,10) -- (0,10);
        \draw [<->,thick] (-1.5,0) -- (-1.5,0 |- 0,10) node [above, rotate=90, midway] {\tiny $N_{\textit{RB}}$};

        \draw[dashed, thin] (-1.5,11) -- (-1.5,13.5);
        \draw[dashed, thin] (8.5,11) -- (8.5,13.5);
        \draw[<->,thick] (-1.5,13.5) -- (8.5,13.5) node [above, midway] {\tiny frame = $10$~ms};

        \draw[dashed, thin] (1.5,11) -- (1.5,12.75);
        \draw[<->,thick] (-1.5,12.75) -- (1.5,12.75) node [above, midway] {\tiny subframe = $1$~ms};

        \draw[dashed, thin] (0,11) -- (0,12);
        \draw[<->,thick] (-1.5,12) -- (0,12) node [above, midway] {\tiny slot};

        \draw[very thin] (-1.5,11) rectangle (0,11.5) node [midway] {\tiny \#0 - 0};
        \draw[very thin] (0,11) rectangle (1.5,11.5) node [midway] {\tiny \#0 - 1};
        \draw[very thin] (1.5,11) rectangle (3,11.5) node [midway] {\tiny \#1 - 0};
        \draw[dotted, very thick] (3.75,11.25) -- (4.75,11.25);
        \draw[very thin] (5.5,11) rectangle (7,11.5) node [midway] {\tiny \#9 -0};
        \draw[very thin] (7,11) rectangle (8.5,11.5) node [midway] {\tiny \#9 -1 };

        \draw[thick] (-1.5,11) rectangle (1.5,11.5) node [midway] {};
        \draw[thick] (5.5,11) rectangle (8.5,11.5) node [midway] {};
        \draw[thick] (1.5,11) -- (3.2,11);
        \draw[thick] (1.5,11.5) -- (3.2,11.5);
        \draw[thick] (5.3,11) -- (5.5,11);
        \draw[thick] (5.3,11.5) -- (5.5,11.5);

        \draw[dashed, thick] (-1.5,11) -- (0,10);
        \draw[dashed, thick] (0,11) -- (7,10);

        \draw[very thin] (0,0) -- (0.5,0) node [below, midway] {\tiny $l=0$};
        \draw[very thin] (8,0) -- (8,0) node [below, midway] {\tiny $l=N_{\textit{symb}}-1$};
        \draw[very thin] (7,0) -- (7,0.5) node [anchor=west, midway] {\tiny $k=0$};
        \draw[very thin] (7,9.5) -- (7,10) node [anchor=west, midway] {\tiny $k=N_{\textit{RB}}\cdot N_{\textit{sc}}^{\textit{RB}} -1$};

    \end{tikzpicture}
    \caption{\footnotesize Radio frame and resource grid with numerology $\mu = 1$. In black, the resource elements occupied by the RUN-O-RAN configured SRS.}
    \label{fig:resource_grid}
\end{figure}
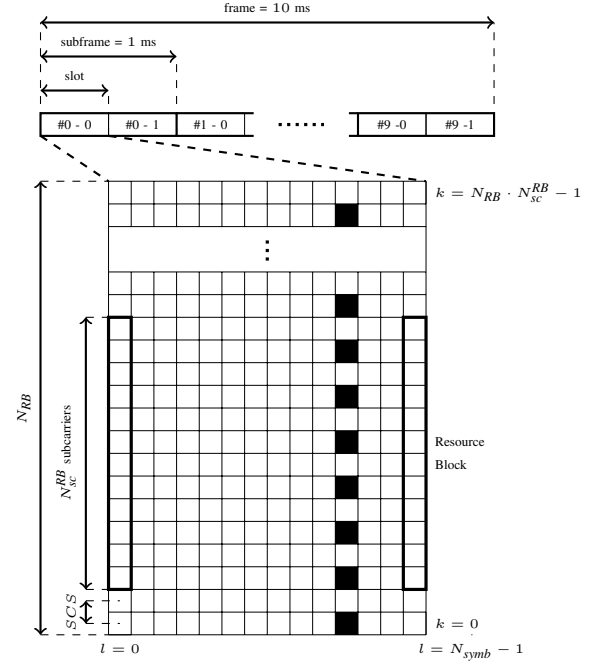

%% file: Content/figures/flowcharts/timing_advance.tex
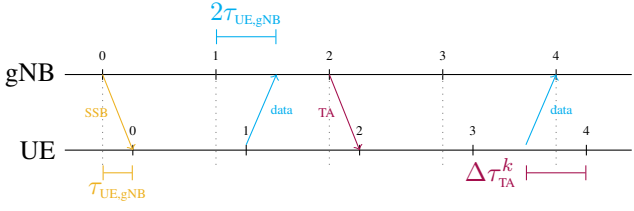
\begin{figure}
\centering
\begin{tikzpicture}[x=1cm,y=1cm]

\def\Tau{0.4}

\node[anchor=east] at (-0.5,0) {gNB};
\draw (-0.5,0) -- (7,0);

\foreach \x in {0,1,2,3,4}{
    \draw (\x*1.5,0.07) -- (\x*1.5,-0.07);
    \node[above] at (\x*1.5,0.07) {\tiny \x};
    \draw[dotted, gray] (\x*1.5,0) -- (\x*1.5,-1.2);
}

\node[anchor=east] at (-0.5,-1) {UE};
\draw (-0.5,-1) -- (7,-1);

\foreach \x in {0,1,2,3,4}{
    \draw ({\x*1.5+\Tau},-1+0.07) -- ({\x*1.5+\Tau},-1-0.07);
    \node[above] at (\x*1.5+\Tau,-1+0.07) {\tiny \x};
}

\draw[->, arancio] (0,0) -- (\Tau,-1)
node[midway,left] {\tiny SSB};

\draw[|-|, arancio] (0,-1.3) -- (\Tau,-1.3)
node[midway,below] {$\tau_{\text{\tiny{UE,gNB}}}$};

\draw[|-|, cyan] (1.5,0.5) -- (1.5+2*\Tau,0.5)
node[midway,above] {$2\tau_{\text{\tiny{UE,gNB}}}$};


\draw[|-|, roxxo] (6-\Tau,-1.3) node[left] {$\Delta \tau_{\text{\tiny{TA}}}^k$}  -- (6+\Tau,-1.3);

\draw[->, roxxo] (3,0) -- (3+\Tau,-1)
node[midway,left] {\tiny TA};

\draw[->, cyan] (1.5+\Tau,-1+0.07) -- (1.5+2*\Tau,0)
node[midway,right,yshift=0pt,xshift=0pt] {\tiny data};

\draw[->, cyan] (6-\Tau,-1+0.07) -- (6,0)
node[midway,right] {\tiny data};

\end{tikzpicture}

\caption{\footnotesize Propagation delay and timing advance correction. The UE downlink timeline is delayed by the propagation time $\tau$ with respect to the gNB reference. After receiving the timing advance command, the UE advances its uplink transmission to align with the gNB timing.}
\label{fig:ta_complete}
\end{figure}

%% file: Content/methodONE.tex
\section{RUN-O-RAN Framework}
\label{sc: run-o-ran}%
RUN-O-RAN turns standard uplink \glspl{srs} transmissions into cooperative localization measurements by allowing serving and non-serving \glspl{gnb} to process the same \gls{ue} waveform. The key obstacle is architectural. In a conventional \gls{ran}, only the serving \gls{gnb} has access to the scheduling decision, resource allocation, and reference sequence associated with a given \gls{ue}. A neighboring \gls{gnb} may physically receive the same \gls{srs}, but it cannot identify the relevant resource elements or perform coherent correlation without this information. This architectural limitation prevents cooperative uplink localization in legacy deployments.

\input{Content/figures/flowcharts/trilateration_basics}
RUN-O-RAN removes this barrier through a centralized xApp deployed in the \gls{nrtric}. The xApp agrees upon the \gls{srs} configuration with the serving anchor, distributes the parameters required for \gls{srs} retrieval to the secondary anchors, and collects their signal observations through the E2 interface. The resulting control and telemetry workflow enables multiple \glspl{gnb} to derive cooperative ranging measurements from the same uplink transmission. Since the entire procedure relies on standard \gls{srs} signaling and is executed within the network, no modification of the chipset or protocol stack on the \gls{cots} \gls{ue} is required. 

The framework separates the localization problem into three functional layers. The first layer is architectural: O-RAN provides the control and telemetry path required to coordinate serving and secondary \glspl{gnb}. The second layer is signal-processing oriented: cooperative \gls{srs} reception, correlation-based \gls{toa} extraction, \gls{ta} compensation, and drift-correction transform uplink reference signals into corrected range estimates. The third layer is algorithmic: trilateration and temporal filtering convert these ranges into a continuous estimate of the \gls{ue} position. The separation between the second and third layers makes RUN-O-RAN extensible, as improvements in anchor selection or tracking can be introduced without changing the underlying coordination mechanism.

Figure~\ref{fig: trilat basics} shows the considered system model. A \gls{ue} is connected to the master, denoted by \textit{gNB}$_m$, and lies within the reception range of at least two secondary anchors, labeled as \textit{gNB}$_s$. More generally, any participating anchor is denoted by \textit{gNB}$_i$, independently of whether it serves the \gls{ue}.

\subsection{O-RAN-Enabled Cooperative Localization Architecture}
\label{sbs: oran architecture}
\input{Content/figures/flowcharts/ORANarchitecture}

Figure~\ref{fig: testbed_emulators} shows the \gls{oran} architecture deployed by RUN-O-RAN. Following the \gls{oran} paradigm, the \gls{gnb} is disaggregated into the \gls{ru}, \gls{du}, and \gls{cu}. This disaggregation is complemented by a programmable control loop that enables interoperable vendor-agnostic control of the \gls{ran}~\cite{oran_polese}: the \gls{nrtric}, which operates within $[10,1000)$~ms and supports near-real-time control via xApps. This makes the \gls{nrtric} the natural location for the RUN-O-RAN control logic.

Communication between the \gls{nrtric} and the underlying \gls{ran} is enabled by the E2 interface. E2 abstracts implementation-specific details and exposes standardized telemetry and control primitives to xApps. Through this interface, xApps can collect measurements, process network-state information, and issue control actions to distributed \gls{ran} nodes while remaining independent of the specific vendor implementation. The E2 interface operation is based on the \gls{e2ap}, and a set of \glspl{e2sm}. \gls{e2ap} manages the association between the RIC and the E2 nodes, whereas \glspl{e2sm} define the semantics and data structures associated with specific monitoring and control functions.%

At the time of deployment, standardized \glspl{e2sm} did not expose all the information required for cooperative uplink positioning. We therefore introduce a dedicated localization-oriented \gls{sm} that extends the E2 interface, a fundamental step in the proof-of-concept development of new xApps~\cite{moro2023open}. The proposed \gls*{sm} introduces semantics that, when extended, pave the way toward \gls{isac} functionalities within \gls{oran}, laying the foundation for a new generation of 6G-oriented \gls{oran} applications.

The \gls*{sm} has been defined and implemented using Protocol Buffers, a language-agnostic, platform-independent format developed by Google LLC to efficiently serialize structured data~\cite{currier2022protocol}.
In particular, RUN-O-RAN \gls*{sm} defines the payload of the messages exchanged between \glspl*{gnb} and the xApp shown in Fig.~\ref{fig: messages}:
\begin{itemize}
    \item $\mathrm{SRS_{conf}}$: reproduces the \gls{3gpp} standard \textit{SRS-Resource} as described in~\ref{sbs: back-reference signals}.
    \item $\mathrm{SRS_{param}}$ contains the subset of $\mathrm{SRS_{conf}}$ parameters required by neighbouring \glspl*{gnb} to identify the time--frequency resources on which the target \gls{ue} transmits its uplink \glspl{srs}.
    \item $\mathrm{SRS}$: efficiently encodes the received baseband \gls{srs} samples together with the hashed \acrlong{crnti} of the associated \gls*{ue} ($\mathrm{UE}_{ID}$), the reception timestamp, and the measured \gls*{snr}.
    \item $\mathrm{TA}_k$: conveys the \gls{tac} value measured by the master \gls*{gnb}, together with the corresponding $\mathrm{UE}_{ID}$ and timestamp.
\end{itemize}

The reader can find further details on GitHub\footnote{ \url{https://github.com/viola-bernazzoli/RUN-O-RAN}}, where the RUN-O-RAN \gls{sm} and a patch implementing the required E2-node extensions were made publicly available.

\subsection{RUN-O-RAN Workflow}
\label{sbs: run-o-ran-workflow}
\input{Content/figures/flowcharts/xAppMex}
To convert a standard uplink reference-signal transmission into a cooperative multi-anchor localization system, we designed a sequence of control and indication messages exchanged between the xApp and the participating \glspl{gnb}, as depicted in Fig.~\ref{fig: messages}. The proposed workflow builds on the E2 procedures and service-model abstractions defined by O-RAN~\cite{oran_e2gap_v410, oran_e2ap_v400,oran_e2sm_v400}. For clarity, the figure omits the E2 association establishment, which follows the standard procedures~\cite{oran_e2gap_v410, oran_e2ap_v400}, and we omit the onboarding, deployment, and registration of the xApp, which are also standard-compliant.
The startup phase is executed once when RUN-O-RAN is instantiated. The subsequent procedures are executed for each newly attached \gls{ue}, whereas the \gls{ta}-update phase is repeated whenever the \gls{tac} procedure is triggered.

\vspace{7pt}
\subsubsection{Startup}
\label{sbs: xapp-startup}

During startup, RUN-O-RAN is deployed in the \gls{nrtric} and requests the identifiers, coordinates, and neighbor relations of the E2-connected \glspl{gnb}. This information defines the set and geometry of the candidate localization anchors. Depending on the deployment, anchor coordinates may be obtained from a network topology database or reported by the \glspl{gnb} through E2 indications. 

Once the infrastructure information has been retrieved, RUN-O-RAN sends control messages $\mathrm{C\{SRS_{conf}\}}$ to the participating \glspl{gnb} to configure the \gls{srs} parameters required by the localization service. This configuration must balance positioning accuracy and radio-resource occupancy. Wider-bandwidth \gls{srs} transmissions and lower periodicity value improve the temporal and spatial resolution of the range estimates, but they also consume more physical resources\footnote{To expose this trade-off to the operator, we define an accuracy--occupancy parameter $\rho_{\mathrm{ao}}\in[0,1]$. The parameter reflects the amount of \glspl{re} occupied by the \gls{srs} in $180$~ms by a single \gls{ue}, and is normalized by the \gls{ue} capabilities. $\rho_{\mathrm{ao}}=0$ and $\rho_{\mathrm{ao}}=1$ represent minimum and maximum occupancy respectively.}.

RUN-O-RAN then subscribes to notifications associated with newly attached \glspl{ue} with $\mathrm{S_{req}\{UE\}}$. The \gls{gnb} reporting the attachment is selected as \textit{gNB}$_m$, while candidate non-serving anchors are obtained from the serving cell's neighbor relations and from the E2-connected nodes capable of monitoring the target carrier. RUN-O-RAN then retains only the candidates that successfully detect the target \gls{ue}'s \gls{srs}.

\vspace{7pt}
\subsubsection{UE Parameter Negotiation}
\label{sbbs: xapp-first}

When a \gls{ue} attaches to the network, \textit{gNB}$_m$ retrieves the \gls{ue} capability information and negotiates an \gls{srs} configuration that is both compatible with the device and suitable for localization. The selected parameters follow the recommendation issued by RUN-O-RAN whenever they are supported by the \gls{ue}; otherwise, \textit{gNB}$_m$ selects the closest feasible configuration.

After each successful random access, represented by the \gls{ra} block in Fig.~\ref{fig: messages}, \textit{gNB}$_m$ sends an indication message $\mathrm{I_{res}\{UE\}}$ to RUN-O-RAN containing: a confidentiality-preserving \gls{ue} identifier $\mathrm{UE_{ID}}$, obtained by hashing the \acrlong{crnti}; the negotiated \gls{srs} configuration $\mathrm{SRS_{conf}}$, which may differ from the requested configuration depending on \gls{ue} capabilities; and the $\mathrm{TA_{RA}}$, namely the first \gls{tac} issued by \textit{gNB}$_m$, later used for timing compensation. These elements identify the time--frequency resources assigned to the \gls{srs} and provide the parameters required to reconstruct the corresponding reference sequence.

RUN-O-RAN then distributes the scheduling and resource-allocation information to the selected non-serving anchors trough a control message $\mathrm{C\{SRS_{param}\}}$. Each anchor monitors the assigned resource elements and reports the corresponding received samples to the xApp, which subscribed to the information through $\mathrm{S_{req}\{SRS\}}$. The xApp obtains the sequence-generation information from \textit{gNB}$_m$ and processes the observations collected by all participating anchors.

\vspace{7pt}
\subsubsection{UE Calibration}
\label{sbbs: xapp-calibration}

The calibration phase is executed for each newly attached \gls{ue}. Its purpose is to initialize the ranging process and estimate the device-dependent clock-drift components, as described in Sec.~\ref{sbbs: second-drift}. In commercial devices, the uplink transmission timing is affected by oscillator drift, chipset-specific corrections, and internal timing-control mechanisms. These effects introduce slowly varying biases in the measured propagation delays and, consequently, systematic errors in the estimated ranges. Since their behavior depends on the \gls{ue} hardware and operating state, the calibration cannot be reused across different devices, but can be reused for different attachment sessions.

During this phase, each participating \textit{gNB}$_i$ reports the received uplink \gls{srs} sequence $s_{gen}$ to the xApp through E2 indications. The xApp centrally estimates the \gls{ue}--\textit{gNB}$_i$ ranges according to Sec.~\ref{sbbs: second-toa} and compensates for timing realignments introduced by \gls{ta} updates, as described in Sec.~\ref{sbbs: second-synch}. 

Once the drift parameters have been estimated, RUN-O-RAN computes the initial \gls{ue} position through trilateration and enters tracking mode. Subsequent \gls{srs} observations are collected by the participating anchors and processed by the xApp to update the position estimate over time.

\subsection{Positioning Pipeline}
\label{sbs: positioning-pipeline}
The cooperative reception procedure provides each participating anchor with a copy of the \gls{srs} transmitted by the target \gls{ue}. However, the delay extracted from this signal cannot be used directly as a geometric range. It contains not only the propagation time between the \gls{ue} and the receiving anchor, but also the effects of uplink synchronization, timing-advance updates, residual inter-\gls{gnb} clock offsets, and chipset-dependent timing corrections. The purpose of the RUN-O-RAN positioning pipeline is therefore to transform raw correlation delays into a corrected distance estimate suitable for trilateration.


\vspace{7pt}
\subsubsection{Signal Model and ToA Extraction}
\label{sbbs: second-toa}

Upon reception of the first \gls{srs} sequence from each anchor, the positioning pipeline is initialized. Since the \gls{ue} uplink timing is aligned with the serving \textit{gNB}$_m$, the delay observed at the master anchor corresponds to a round-trip propagation term proportional to $2d_{\mathrm{UE},m}/c$, where $d_{\mathrm{UE},i}$ is the distance between the \gls{ue} and the $i$-th anchor, and $c$ is the signal propagation speed. At a secondary anchor, instead, the downlink timing reference is still determined by \textit{gNB}$_m$, whereas the uplink signal is received by \textit{gNB}$_s$. Therefore, the measured delay contains the composite propagation term $(d_{\mathrm{UE},m}+d_{\mathrm{UE},s})/c$.

Let $s_{\mathrm{gen}}$ denote the generated \gls{srs} sequence, known by the \gls{ue}, \textit{gNB}$_m$, and RUN-O-RAN after the attachment. Let $s_{i,\mathrm{rx}}$ be the sequence received at the $i$-th anchor and sampled at frequency $f_s$. We assume that all participating anchors operate over the same \gls{srs} bandwidth and therefore use the same sampling frequency. The received signal can be modeled as
\begin{equation}
\label{eq: signal model}
s_{i,\mathrm{rx}}[n] =
\alpha_i[n]s_{\mathrm{gen}}[n-\widetilde{m}_i] + \eta_i[n],
\quad n \in [0,N_{SC}^{SRS}],
\end{equation}
where $n$ is the sample index, $\alpha_i[n]$ is the channel attenuation term between the \gls{ue} and \textit{gNB}$_i$, $\widetilde{m}_i$ is the discrete reception delay, $\eta_i[n]$ is additive noise, and $N_{SC}^{SRS} = m_{SRS,b} \cdot N_{sc}^{RB}/K_{TC}$ is the length of the \gls{srs} sequence.

RUN-O-RAN estimates the discrete channel impulse response by cross-correlating the received signal with the known reference:
\begin{equation}
\hat{h}_i[n] =
\mathcal{R}_{s_{i,\mathrm{rx}},s_{\mathrm{gen}}}[n],
\end{equation}
where $\mathcal{R}$ 
denotes the cross-correlation operator. Under the correlation properties of \gls{srs} sequences, $\hat{h}_i[n]$ approximates a scaled and shifted version of the autocorrelation of the transmitted waveform. Peaks of $\hat{h}_i[n]$ therefore correspond to sampled propagation paths between the \gls{ue} and the receiving anchor.

In many ranging systems, the \gls{toa} is estimated by selecting the strongest correlation peak. This criterion is reliable only when the strongest path also corresponds to the first arriving path. In multipath environments, however, the highest-power component may be a reflected path and may therefore arrive later than the direct or shortest path. To mitigate this effect, RUN-O-RAN first identifies the set of significant peaks and the delay estimate is then obtained as the earliest significant peak:
\begin{equation}
    \hat{m}_i = \arg\min_{n\in\mathcal{P}_i} n, \quad \text{for} \; \mathcal{P}_i = \{ n : |\hat{h}_i[n]|^2 \geq \gamma_i \},
\end{equation}
where $\gamma_i$ is a detection threshold selected according to the noise floor and correlation sidelobe level; this choice favors the first detectable path and therefore estimates the true \gls{los} delay in \gls{los} multipath scenarios, while minimizing positive bias in \gls{nlos} conditions.

The corresponding distance estimate is obtained after compensating for the master-anchor reference delay. For the $i$-th anchor, the raw range estimate can be written as
\begin{equation}
\label{eq: distance_formula_compact}
\hat{d}_{\mathrm{UE},i}
=
\Delta d
\left(
\hat{m}_i - \frac{1}{2}\hat{m}_m
\right), \quad \Delta d =
\frac{c}{n_{\mathrm{FFT}}\cdot \mathrm{SCS}},
\end{equation}
where $\hat{m}_m$ is the delay index estimated at the master \textit{gNB}$_m$, and $\Delta d$ is the distance resolution of the correlation domain. This resolution is determined by the \gls{srs} bandwidth and sampling configuration, where $n_{\mathrm{FFT}}$ is the Fourier-transform size used to discretize the \gls{srs} in the frequency domain, and $\mathrm{SCS}$ is the subcarrier spacing. Increasing the \gls{srs} bandwidth improves the granularity of the correlation space and therefore the achievable ranging accuracy, at the cost of higher radio-resource consumption.

The same procedure is repeated independently for all participating anchors. The resulting raw measurements are then passed to the \gls{ta} and drift compensation stages described below.
\label{sbs: ranging}

\begin{figure*}[t]
\centering
\begin{subfigure}{0.32\textwidth}
    \centering
    \includegraphics[width=\linewidth]{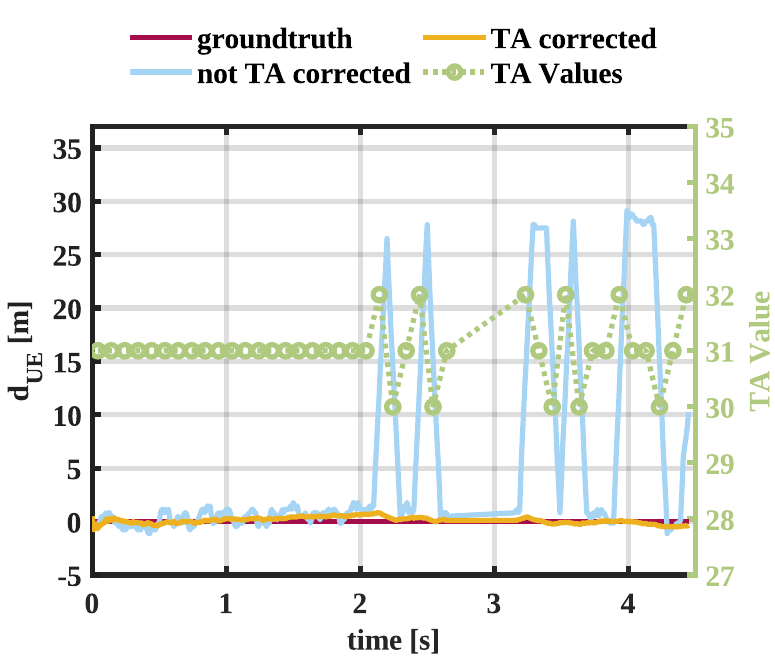}
    \caption{\footnotesize \gls*{ta}-corrected}
    \label{fig: exp_static_ta}
\end{subfigure}
\hfill
\begin{subfigure}{0.32\textwidth}
    \centering
    \includegraphics[width=\linewidth]{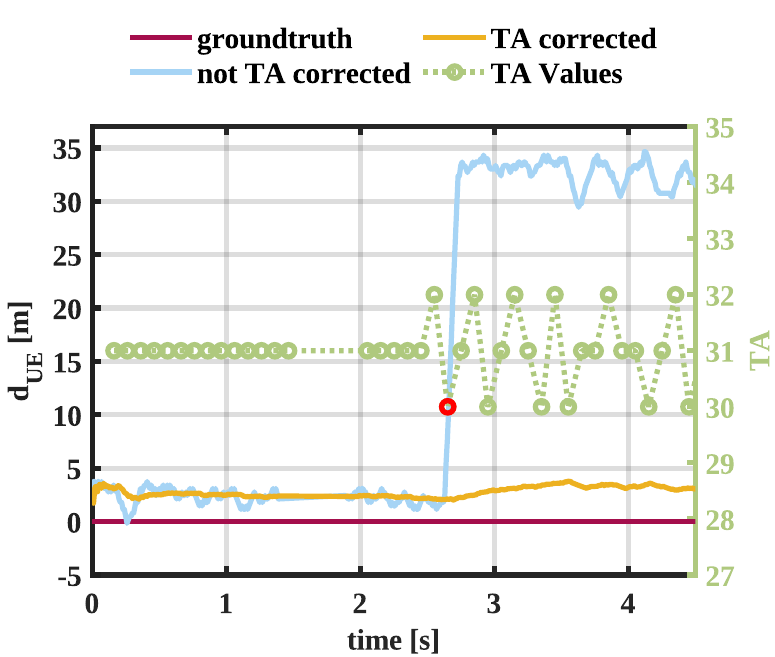} 
    \caption{\footnotesize lost-\gls*{ta}-corrected}
    \label{fig: exp_static_ta_lost}
\end{subfigure}
\hfill
\begin{subfigure}{0.32\textwidth}
    \centering
    \includegraphics[width=\linewidth]{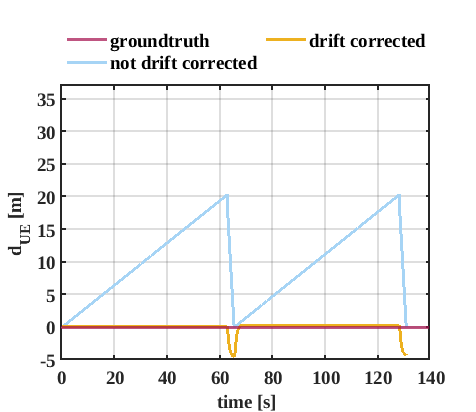} 
    \caption{\footnotesize Drift-corrected}
    \label{fig: exp_static_drift}
\end{subfigure}
\caption{\footnotesize Correction steps of RUN-O-RAN. The red line represents the ground truth, and the azure and orange ones represent $\hat{d}_{UE}$ before and after the correction, respectively.}
\label{fig:exp_steps_of_the_way}
\end{figure*}
\vspace{7pt}
\subsubsection{Timing Advance Compensation}
\label{sbbs: second-synch}
As discussed in Sec.~\ref{sbs: back-synchronization}, uplink transmissions are periodically realigned to the resource grid of \textit{gNB}$_m$ through the \gls{ta} mechanism. Accurate tracking of \glspl{tac} is therefore essential, since each \gls{ta} update modifies the effective transmission timing of the \gls{ue}. RUN-O-RAN subscribes to these updates during the parameter-negotiation phase. Whenever \textit{gNB}$_m$ sends a \gls{tac} to the tracked \gls{ue}, it also forwards the corresponding information to the xApp.

This allows RUN-O-RAN to maintain an updated record of the cumulative timing advance applied to the \gls{ue} and to compensate for its effect on the estimated ranges. The \gls{ta}-corrected distance is computed as:
\begin{equation}
\hat{d}_{\mathrm{UE},i}^{\,\mathrm{TA\text{-}Corr.}}
=
\hat{d}_{\mathrm{UE},i}
-
\frac{c\cdot \tau_{\text{\tiny{TA}}}^k}{2},
\end{equation}
where $\tau_{\text{\tiny{UE}}}^k$ is derived with eq.~\ref{eq: tau}. Figure~\ref{fig: exp_static_ta} shows the effect of \gls{ta} compensation on the range estimate. While the uncorrected estimate $\hat{d}_{\mathrm{UE},m}$ exhibits large deviations following \gls{tac} updates, the corrected estimate $\hat{d}_{\mathrm{UE},m}^{\,\mathrm{TA\text{-}Corr.}}$ remains close to the ground-truth distance.

A practical difficulty arises from the absence of an explicit \gls{tac} feedback mechanism. As a result, we have no \gls{ue} reception confirmation. If a \gls{tac} is lost or corrupted, the xApp's estimate of the \gls{ue} transmission timing becomes inconsistent with the actual device behavior, which can invalidate both range and position estimates. RUN-O-RAN therefore detects missed \gls{ta} commands by checking the plausibility of consecutive range updates. If a corrected range variation implies a physically unrealistic apparent speed, for example for $\mu=1$ the threshold $\textit{thr}=\Delta \tau_{\mathrm{UE}}^k \big{|}_{\mathrm{TA}_k=32}\cdot \nicefrac{c}{\mathrm{srs\_periodicity}} \approx 139$~m/s, the corresponding \gls{tac} is flagged as potentially unacknowledged, and the cumulative timing-advance record is adjusted accordingly. An example is shown in Fig.~\ref{fig: exp_static_ta_lost}, where the red marker denotes a \gls{tac} observed by the xApp but not by the \gls{ue}. Including this missed command in the cumulative \gls{ta} introduces a persistent bias in $\hat{d}_{\mathrm{UE},m}^{,\mathrm{TA\text{-}Corr.}}$. Once detected by RUN-O-RUN, however, the timing-advance record is corrected, yielding the orange estimate.

\vspace{7pt}
\subsubsection{Clock-Drift Calibration and Correction}
\label{sbbs: second-drift}

Even after \gls{ta} compensation, accurate localization requires tracking residual timing bias. In commercial devices, the \gls{ue} transmission timing is affected by clock drift, chipset-dependent corrections, and internal timing-control mechanisms. These effects introduce slowly varying biases that appear as systematic errors in the measured ranges.

According to \gls{3gpp} specifications, the maximum \gls{ue} transmission timing error $T_e^{\max}$ must remain below the \gls{ta} discretization step~\cite{3gpp-ts-38.533-v15.0.0}. To satisfy this requirement, the \gls{ue} continuously estimates its timing error $T_e$ using downlink reference signals. When the error exceeds the allowed bound, the \gls{ue} gradually adjusts its transmission timing through small consecutive corrections.

Empirical observations show that this behavior produces a repeatable sawtooth timing-drift pattern~\cite{eurecom_drift}. The pattern alternates between coasting phases, during which clock bias accumulates approximately linearly, and corrective phases, during which the \gls{ue} progressively compensates the accumulated error, as the blue line shows in Fig.~\ref{fig: exp_static_drift}. For a given \gls{ue} model, the slopes of these phases are sufficiently consistent across experiments to be learned and later compensated. RUN-O-RAN exploits this regularity by estimating the characteristic drift slopes during a calibration stage and subtracting their contribution from subsequent range estimates.

\textit{Timing-bias calibration.}
When a \gls{ue} attaches to \textit{gNB}$_m$, RUN-O-RAN initiates a one-time calibration procedure before enabling reliable tracking. A lightweight \gls{kf}, evolving according to a random-walk model, jointly tracks the apparent distance, timing bias, and bias derivative. By analyzing the variance of the second derivative of the bias estimate, the xApp identifies intervals in which the \gls{ue} can be considered static. During these intervals, RUN-O-RAN estimates two characteristic slopes: a positive slope $\hat{\delta}_+$ associated with the coasting phase, and a negative slope $\hat{\delta}_-$ associated with the corrective phase.

\textit{Timing-bias correction.}
Once calibration is complete, RUN-O-RAN compensates timing drift in real time. The xApp identifies the beginning of each coasting or corrective phase by monitoring the sign and evolution of the first and second derivatives of the estimated timing bias. Depending on the current phase, it selects either $\hat{\delta}_+$ or $\hat{\delta}_-$ and applies the corresponding correction to each anchor-specific range:
\begin{equation}
\label{eq: driftone}
\tilde{d}_{\mathrm{UE},i}
=
\hat{d}_{\mathrm{UE},i}^{\mathrm{TA\text{-}Corr.}}
-
c
\hat{\delta}_{\nicefrac{+}{-}}
\Delta t_{\mathrm{sync}},
\end{equation}
where $\hat{\delta}_{\nicefrac{+}{-}}$ denotes the active drift slope, selected according to the current sawtooth phase, 
and $\Delta t_{\mathrm{sync}}$ is the time elapsed since the beginning of the current coasting or corrective interval. This correction removes the dominant deterministic timing bias, leaving only residual stochastic fluctuations to be handled by the final position estimator.

\subsection{Position Estimation}
\label{sbs: position-estimation}

Once the ranging pipeline has produced the corrected anchor-wise measurements, localization can be cast as a geometric estimation problem. At SRS instant $k$, RUN-O-RAN observes the mixed measurement vector
\begin{equation}
\label{eq:position-measurement}
\mathbf{z}_k =
\begin{bmatrix}
\tilde{d}_{\mathrm{UE},m}^{\, k} &
\tilde{d}_{\mathrm{UE},s1}^{\, k} -  \tilde{d}_{\mathrm{UE},m}^{\, k} &
\tilde{d}_{\mathrm{UE},s2}^{\, k} - \tilde{d}_{\mathrm{UE},m}^{\, k}
\end{bmatrix}^{\mathsf T},
\end{equation}
where $\tilde{d}_{\mathrm{UE},m}^{\, k}$ is the corrected absolute range from the master anchor, and $\tilde{d}_{\mathrm{UE},s1}^{\, k} - \tilde{d}_{\mathrm{UE},m}^{\, k}$ and $\tilde{d}_{\mathrm{UE},s2}^{\, k} - \tilde{d}_{\mathrm{UE},m}^{\, k}$ are the corrected differential ranges of the two secondary anchors with respect to the master reference.

Let $\mathbf{p}_k = [x_k, y_k]^{\mathsf T}$ denote the UE position and let $\mathbf{u}_i = [x_i, y_i]^{\mathsf T}$ be the known position of anchor $i$. Using the master anchor $m$ and the two secondaries $s_1$ and $s_2$, the corresponding nonlinear measurement model is
\begin{equation}
\label{eq:position-model}
\mathbf{h}(\mathbf{p}) =
\begin{bmatrix}
\|\mathbf{p} - \mathbf{u}_m\|_2 \\
\|\mathbf{p} - \mathbf{u}_{s_1}\|_2 - \|\mathbf{p} - \mathbf{u}_m\|_2 \\
\|\mathbf{p} - \mathbf{u}_{s_2}\|_2 - \|\mathbf{p} - \mathbf{u}_m\|_2
\end{bmatrix}.
\end{equation}

Measurement reliability is encoded by
\begin{equation}
\label{eq:position-weights}
\mathbf{R} = \operatorname{diag}\!\left(\sigma_m^2,\sigma_t^2,\sigma_t^2\right),
\qquad
\mathbf{W} = \mathbf{R}^{-1/2},
\end{equation}
where $\sigma_m$ and $\sigma_t$ are the standard deviations assigned to the master absolute range and to the secondary anchors' differential measurements, respectively. This common formulation is used by both estimators below.
\vspace{7pt}
\subsubsection{Weighted Nonlinear Least Squares}
\label{subsubsec:position-nls}

the algorithm computes the position independently at each \gls{srs} instant by minimizing the weighted residual between the observation vector in~\eqref{eq:position-measurement} and the model in~\eqref{eq:position-model}:
\begin{equation}
\label{eq:position-nls}
\hat{\mathbf{p}}_k^{\mathrm{NLS}}
=
\arg\min_{\mathbf{p}\in\mathbb{R}^2}
\rho\!\left(
\left\|
\mathbf{W}\bigl(\mathbf{z}_k - \mathbf{h}(\mathbf{p})\bigr)
\right\|_2^2
\right),
\end{equation}
where $\rho(\cdot)$ is a robust penalty, such as soft-$L_1$, used to reduce the influence of isolated outliers. In practice, the solver is initialized with the previous estimate $\hat{\mathbf{p}}_{k-1}^{\mathrm{NLS}}$ when available, and with the anchor centroid otherwise. This yields a compact geometric estimate at each \gls{srs} occasion without introducing an explicit mobility model.
\vspace{7pt}
\subsubsection{Extended Kalman-Based Tracking}
\label{subsubsec:position-kf}

To exploit temporal continuity, the tracker uses the same measurement geometry within a constant-velocity \gls{ekf}. The state is described by
\begin{equation}
\label{eq:position-state}
\mathbf{x}_k =
\begin{bmatrix}
\mathbf{p}_k^{\mathsf T} & \mathbf{v}_k^{\mathsf T}
\end{bmatrix}^{\mathsf T}
=
\begin{bmatrix}
x_k & y_k & \dot x_k & \dot y_k
\end{bmatrix}^{\mathsf T},
\end{equation}
with transition model
\begin{equation}
\label{eq:position-transition}
\mathbf{x}_{k|k-1} = \mathbf{F}_k \mathbf{x}_{k-1|k-1},
\qquad
\mathbf{F}_k =
\begin{bmatrix}
\mathbf{I}_2 & \Delta t_k \mathbf{I}_2 \\
\mathbf{0}_2 & \mathbf{I}_2
\end{bmatrix}.
\end{equation}

Assuming white acceleration noise with standard deviation $\sigma_a$, the process covariance is
\begin{equation}
\label{eq:position-process-noise}
\mathbf{Q}_k = \sigma_a^2
\begin{bmatrix}
\Delta t_k^4/4 & 0 & \Delta t_k^3/2 & 0 \\
0 & \Delta t_k^4/4 & 0 & \Delta t_k^3/2 \\
\Delta t_k^3/2 & 0 & \Delta t_k^2 & 0 \\
0 & \Delta t_k^3/2 & 0 & \Delta t_k^2
\end{bmatrix}.
\end{equation}

The measurement function is still given by~\eqref{eq:position-model}, so the linearized measurement matrix results in:
\begin{equation}
\label{eq:position-jacobian}
\mathbf{H}_k
=
\left[
\frac{\partial \mathbf{h}}{\partial \mathbf{p}}
\bigg|_{\mathbf{p}=\hat{\mathbf{p}}_{k|k-1}}
\;\;
\mathbf{0}_{3\times 2}
\right].
\end{equation}

The \gls{ekf} is then updated as standard.

\vspace{7pt}
The two estimators therefore differ only in their temporal structure: $\hat{\mathbf{p}}_k^{\mathrm{NLS}}$ uses the current measurement only, whereas $\hat{\mathbf{x}}_k^{\mathrm{EKF}}$ combines the same measurement geometry with a constant-velocity prior.

%% file: Content/figures/flowcharts/trilateration_basics.tex
\begin{figure}[b]
    \centering
    \begin{tikzpicture}[scale=0.5]
    
    \coordinate (gnbm) at (0,0);
    \coordinate (gnbs1) at (4,0);
    \coordinate (gnbs2) at (2,3);
    
    \coordinate (ue) at (2,1);
    
    \def\rone{2.236}   
    \def\rtwo{2.236}   
    \def\rthree{2.000} 
    \definecolor{mydarkgreen}{RGB}{19,42,19}
    
    \draw (gnbm) circle (\rone);
    \draw (gnbs1) circle (\rtwo);
    \draw (gnbs2) circle (\rthree);
    
    \fill (gnbm) circle (2pt);
    \fill (gnbs1) circle (2pt);
    \fill (gnbs2) circle (2pt);
    
    \node[left] at (gnbm) {\textit{gNB}$_m$};
    \node[right] at (gnbs1) {\textit{gNB}$_{s1}$};
    \node[above] at (gnbs2) {\textit{gNB}$_{s2}$};
    
    \draw[thick, cyan] (ue) ++(-0.12,-0.12) -- ++(0.24,0.24);
    \draw[thick, cyan] (ue) ++(-0.12,0.12) -- ++(0.24,-0.24);
    \node[above right, cyan] at (ue) {UE};
    
    \draw[<->, thick] ($(gnbm)+(0.2,0.1)$) -- (1.8,0.9);
    \draw[<-, thick] ($(gnbs1)-(0.2,-0.1)$) -- (2.15,0.9);
    \draw[<-, thick] ($(gnbs2)-(0,0.2)$) -- (2,1.2);
    
    \end{tikzpicture}

\caption{\footnotesize Trilateration geometry used for UE localization. The UE is connected to the serving base station \textit{gNB}$_m$ and is simultaneously listened by two neighboring anchors \textit{gNB}$_{s1}$ and \textit{gNB}$_{s2}$.}
\label{fig: trilat basics}
\end{figure}

%% file: Content/figures/flowcharts/ORANarchitecture.tex
\begin{figure} 
    \centering

    \begin{tikzpicture}[scale = 0.7]
        
        \foreach \i in {0, 3.5, 7}
        {
            \draw[thin] (0,\i) rectangle (3.6,2.7+\i);
            \node[cylinder, draw, minimum width = 1.5cm, minimum height = 0.9cm, aspect = 0.2, shape border rotate = 90] (c) at (-1.5,0.45+\i) {\tiny DATA};
            \draw [dashed, very thin]  (-0.5,0.45+\i) -- (0.2,0.45+\i);
            \draw[very thin, dashed] (0.2,0.2+\i) rectangle (2.2,0.7+\i) node [midway] {\tiny emulated PHY};
            \draw[very thin] (0.1,0.1+\i) rectangle (3.5,0.8+\i);
            \node at (3,0.45+\i) {\tiny O-RU};
            \draw[very thin] (1.2,0.8+\i) -- (1.2,1+\i);
            \draw[very thin] (0.1,1+\i) rectangle (2.3,1.7+\i) node [midway] {\tiny O-DU};
            \draw[very thin] (1.2,1.7+\i) -- (1.2,1.9+\i);
            \draw[very thin] (0.1,1.9+\i) rectangle (2.3,2.6+\i) node [midway] {\tiny O-CU};
            \draw[very thin] (2.5,1+\i) rectangle (3.5,2.6+\i) node [midway, rotate=270] {\tiny E2 Agent};
            \draw[very thin] (2.3,1.35+\i) -- (2.5,1.35+\i);
            \draw[very thin] (2.3,2.25+\i) -- (2.5,2.25+\i);
            \draw[very thin] (3.5,1.8+\i) -- (5,5.3) node [above, midway] {\tiny $E2$};
        }   

        \node at (-1,0.39) {\tiny $_{s2}$};
        \node at (-1,0.39+3.5) {\tiny $_{s1}$};
        \node at (-1,0.39+7) {\tiny $_m$};
        
        \node at (3.3,2.9) {\tiny gNB$_{s2}$};
        \node at (3.3,2.9+3.5) {\tiny gNB$_{s1}$};
        \node at (3.3,2.9+7) {\tiny gNB$_m$};

        \draw[very thin] (5,5.3) -- (5.6,5.3);
        \draw[thin] (5.5,1.4) rectangle (8.1,6.3);
        \node at (6.8,6.5) {\tiny $nRT\textit{-}RIC$};

        \draw[very thin] (5.6,4.5) rectangle (6.6,6.1) node [midway, rotate=270] {\tiny E2 Term};
        \draw[very thin] (6.6,5.3) -- (6.8,5.3) -- (6.8, 4.3);
        
        \filldraw[very thin, roxxo] (5.6,2.5) rectangle (8,4.3);
        \node[roxxo] at (8,4.5) {\tiny RUN-O-RAN};
        \fill[white] (5.7,3.5) rectangle (7.9,4.2);
        \fill[white] (5.7,2.6) rectangle (7.9,3.3);
        \draw[very thin] (5.7,3.5) rectangle (7.9,4.2) node [midway] {\tiny Ranging};
        \draw[very thin] (5.7,2.6) rectangle (7.9,3.3) node [midway] {\tiny Trilateration};
        \draw[very thin] (6.8,3.3) -- (6.8,3.5);

        \draw[very thin, dashed] (5.6,1.6) rectangle (8,2.3) node [midway] {\tiny other xApps...};

    \end{tikzpicture}
    \caption{\footnotesize RUN-O-RAN validation architecture. Synchronized traces collected for the serving anchor (\textit{gNB}$_m$) and the secondary anchors (\textit{gNB}$_{s1}$, \textit{gNB}$_{s2}$) are injected into the emulated PHY layers of three O-RAN-compatible gNB instances. Each node reports its measurements to the near-RT RIC through E2, where the RUN-O-RAN xApp performs ranging and trilateration.}
    \label{fig: testbed_emulators}
\end{figure}

%% file: Content/figures/flowcharts/xAppMex.tex
\begin{figure}[t]
\vspace*{-0.45cm}
\hspace*{-0.3cm}
\resizebox{9cm}{!}{
\begin{adjustbox}{
        trim={0 1.5cm 0 0},
        clip
    }
\small{
\begin{msc}[draw frame=none,draw grid=none, msc keyword=, title top distance=0cm, left environment distance=1.1cm, right environment distance=0.6cm, label distance = 0.5ex]{}
    \declinst[cyan, label distance=0.2cm]{ue}{UE}{target}
    \declinst[instance distance = 1.6cm, label distance=0.2cm]{gnb}{$gNB_{m}$}{master}
    \declinst[instance distance = 1.6cm, label distance=0.2cm, roxxo]{xapp}{xApp}{controller}
    \declinst[instance distance = 1.6cm, label distance=0.2cm]{gnb2}{$gNB_{s}$}{secondary}

    \measure[right,side=left]{\rotatebox[origin=c]{90}{Startup}}{envleft}{envleft}[5] 
    \mess*{}{envright}{envleft}    
    
    \nextlevel[1.3]
    \mess{$\mathrm{C\{\mathrm{SRS_{conf}}\}}$}{xapp}{gnb}
    \nextlevel[1.3]
    \mess{$\mathrm{S_{req}\{UE}\}$}{xapp}{gnb}
    
    \nextlevel
    \mess*{}{envright}{envleft}
    \measure[right,side=left]{\rotatebox[origin=c]{90}{Parameter Negotiation}}{envleft}{envleft}[17] 

    \nextlevel[0.5]
    \inlinestart[right inline overlap=0.2cm, left inline overlap=0.7cm]{rar}{RA}{ue}{gnb}
    \nextlevel
    \mess{MSG1}{ue}{gnb}
    \nextlevel
    \mess{MSG2}{gnb}{ue}
    \nextlevel
    \mess{MSG3}{ue}{gnb}
    \nextlevel
    \mess{MSG4}{gnb}{ue}
    \nextlevel[0.4]
    \inlineend{rar}

    \nextlevel
    \mess{$\mathrm{I_{res}\{UE\}}$}{gnb}{xapp}
    \mess[label position=below, label distance= 1ex]{\tiny $\mathrm{UE_{ID},SRS_{conf},TA_{RA}}$}{gnb}{xapp} 
    \nextlevel[0.5]
    \mess{$\mathrm{C\{SRS_{param}\}}$}{xapp}{gnb2}
    \nextlevel[1.5]
    \mess{$\mathrm{S_{req}\{SRS\}}$}{xapp}{gnb}
    \mess{$\mathrm{S_{req}\{SRS\}}$}{xapp}{gnb2}

    \nextlevel
    \mess*{}{envright}{envleft}
    \measure[right,side=left]{\rotatebox[origin=c]{90}{Calibration}}{envleft}{envleft}[14]

    \nextlevel[1.5]
    \mess{$s_{gen}$}{ue}{gnb}
    \mess*{$s_{gen}$}{envright}{gnb2}
    \nextlevel[0.4]
    \mess{$\mathrm{I_{res}\{SRS\}}$}{gnb}{xapp}
    \mess{$\mathrm{I_{res}\{SRS\}}$}{gnb2}{xapp}
    \nextlevel[0.5]
    \action*{TA Compensation\textcolor{white}{.}}{xapp}
    \nextlevel[1.5]
    \action*{\textcolor{white}{...}Clock drift est.\textcolor{white}{...}}{xapp}
    \nextlevel[1.5]
    \action*{\textcolor{white}{..}Position update\textcolor{white}{...}}{xapp}
    \nextlevel[2]
    \mess*{}{envright}{envleft}

    \nextlevel[1.5]
    \mess{$s_{gen}$}{ue}{gnb}
    \mess*{$s_{gen}$}{envright}{gnb2}
    \nextlevel[0.4]
    \mess{$\mathrm{I_{res}\{SRS\}}$}{gnb}{xapp}
    \mess{$\mathrm{I_{res}\{SRS\}}$}{gnb2}{xapp}
    \nextlevel[0.5]
    \action*{\textcolor{white}{..}Position update\textcolor{white}{...}}{xapp}
    \nextlevel[1.8]
    \mess{$s_{gen}$}{ue}{gnb}
    \mess*{$s_{gen}$}{envright}{gnb2}
    \nextlevel[0.4]
    \mess{$\mathrm{I_{res}\{SRS\}}$}{gnb}{xapp}
    \mess{$\mathrm{I_{res}\{SRS\}}$}{gnb2}{xapp}
    \nextlevel[0.5]
    \action*{\textcolor{white}{..}Position update\textcolor{white}{...}}{xapp}

    \nextlevel[2]
    \mess*{}{envright}{envleft}
    \measure[right,side=left]{\rotatebox[origin=c]{90}{TA}}{envleft}{envleft}[17]
    \nextlevel[0.5]
    \inlinestart[right inline overlap=0.9cm, left inline overlap=0.95cm]{ta}{TA}{ue}{gnb}
    \nextlevel[0.4]
    \action*{TA Eval.}{gnb}
    \nextlevel[2.5]
    \mess{TA Command}{gnb}{ue}
    \nextlevel[0.5]
    \mess{$\mathrm{I_{res}\{TA_k\}}$}{gnb}{xapp}
    \action*{TA Apply}{ue}
    \nextlevel[0.5]
    \action*{TA Compensation\textcolor{white}{.}}{xapp}
    \nextlevel[1]
    \inlineend{ta}
    \nextlevel[1]
    \mess*{}{envright}{envleft}

    \nextlevel
\end{msc}
}
\end{adjustbox}
}
\caption{\footnotesize Exchange of messages between UE, gNB master, xApp, and secondary gNBs. Notation: E2AP RIC Control Message ($\mathrm{C\{\cdot\}}$), E2AP RIC Subscrption Request ($\mathrm{S_{req}\{\cdot\}}$), E2AP RIC Indication Response ($\mathrm{I_{res}\{\cdot\}}$).}
\label{fig: messages}
\end{figure}

%% file: Content/experimentalSetup.tex
\section{Experimental Setup}
\label{sc: setup}

\begin{figure}[t] 
    \centering
    \begin{subfigure}{\linewidth}
    \centering
        \includegraphics[width=\linewidth]{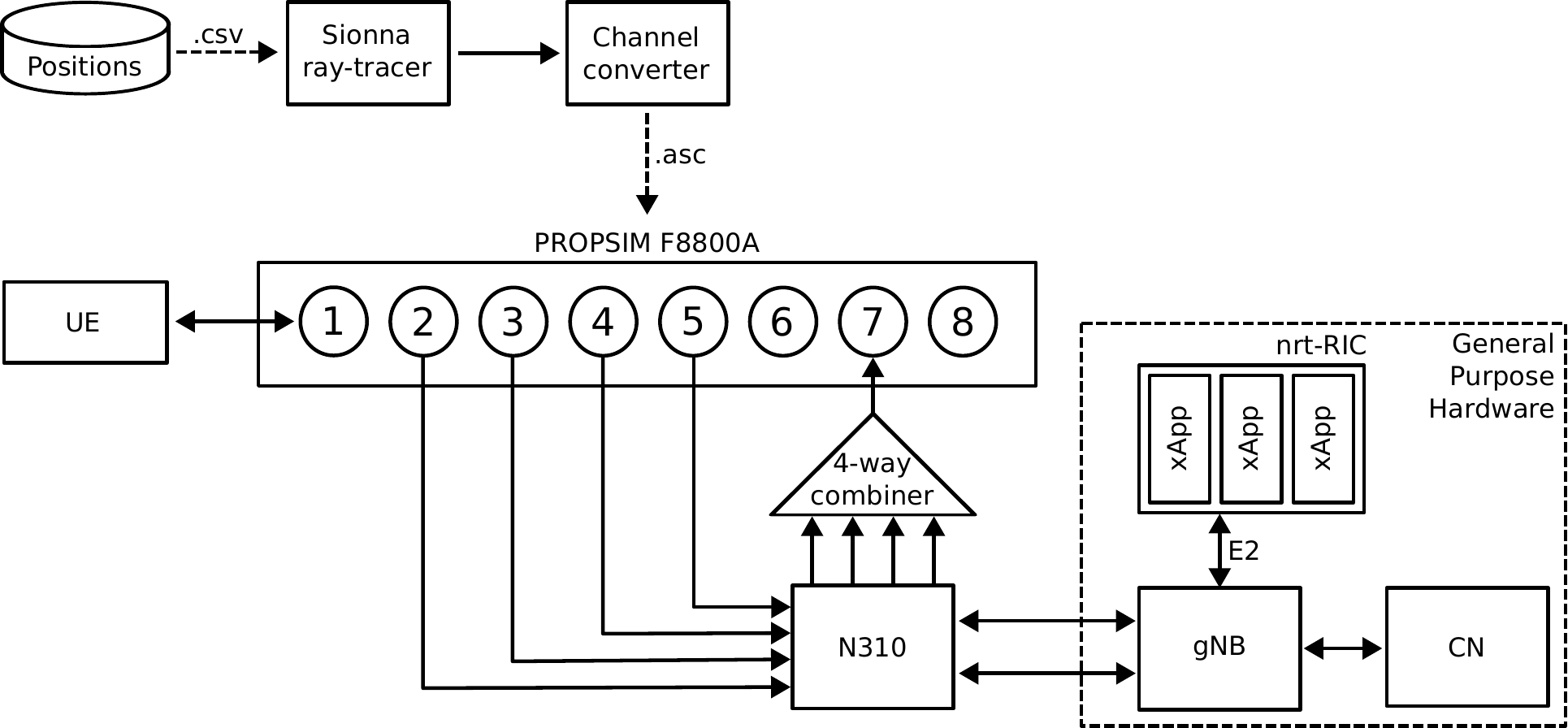}
    \label{fig: real setup}
    \end{subfigure}
    \hfill
    \begin{subfigure}{\linewidth}
    \includegraphics[width=\linewidth]{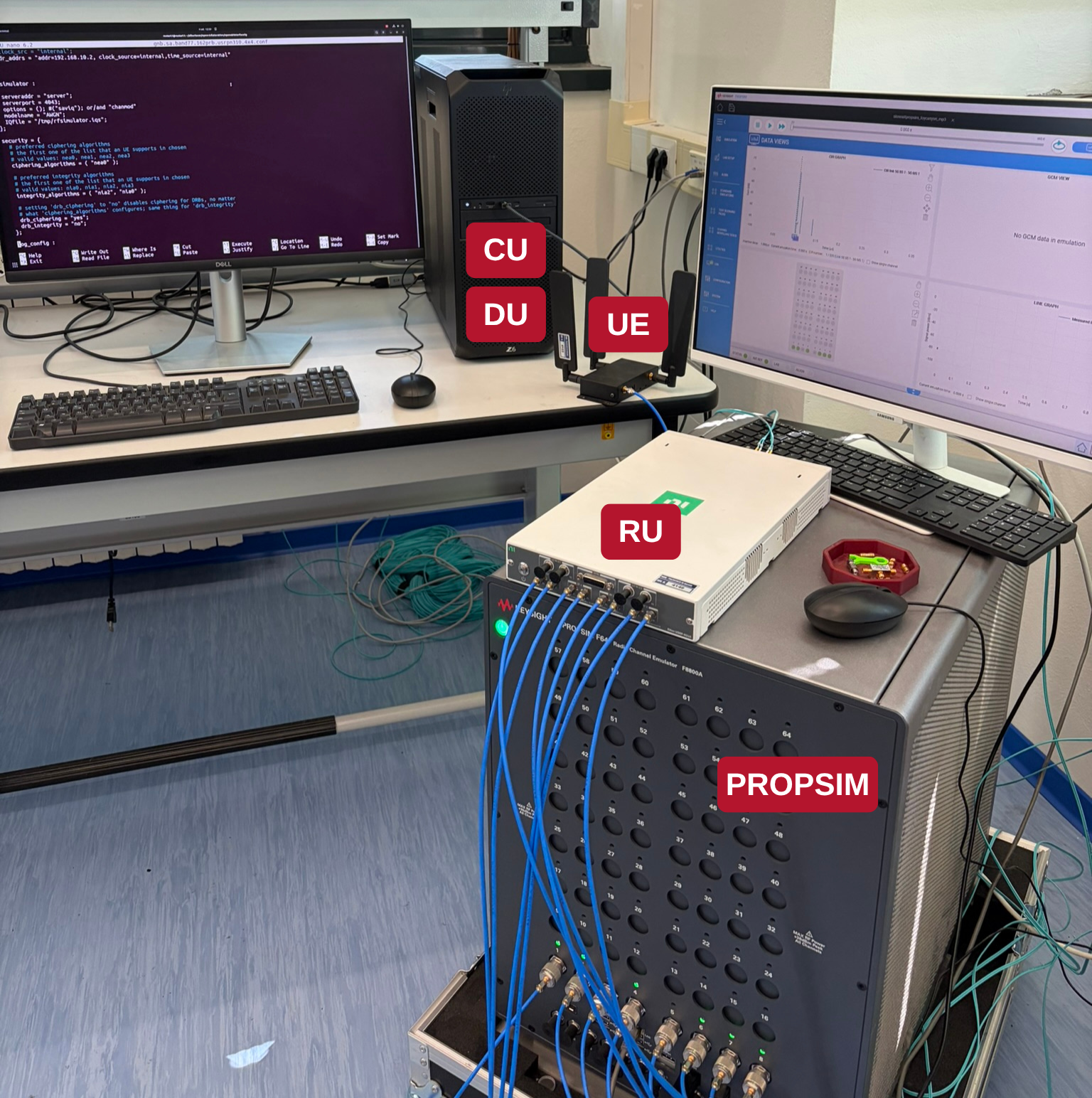}
    \label{fig: siopsim flowchart}
    \end{subfigure}
    \vspace{-0.5cm}
    \caption{\footnotesize Experimental setup used to collect the traces replayed by the e-PHY. Sionna-rt generates the propagation scenario, whose channel realizations are instantiated by the channel emulator. A COTS 5G \gls{ue} is connected to an OAI \gls{gnb} and transmits standard uplink \gls{srs} waveforms, which are simultaneously recorded through three independently configured propagation paths.}
    \label{fig: siopsim}
\end{figure}

We validate RUN-O-RAN in real time through the experimental architecture shown in Fig.~\ref{fig: testbed_emulators}, which was designed to execute the complete RUN-O-RAN workflow under controlled and reproducible propagation conditions. 

The testbed uses three \gls{oai} \glspl{gnb} instances~\cite{openairinterface} connected to an Open5GS \gls{cn}~\cite{open5gs}. We rely on open-source implementations that are easy to enhance by introducing new capabilities. Indeed, we extended the \gls{du} by integrating the RUN-O-RAN \gls{e2sm} and exposing the configuration and measurement information required for cooperative uplink processing~\ref{sbs: oran architecture}. The three anchors establish independent E2 connections with the O-RAN Software Community (SC) \gls{nrtric}, where RUN-O-RAN is deployed. 

A physical deployment with three independent SDR-based \glspl{gnb} would require tight, potentially expensive, inter-node time and frequency synchronization, as commercial \glspl{ran} typically provide. Depending on the synchronization system implemented, residual inter-gNB clock offsets would directly affect the measured arrival times.
To isolate the localization framework from these external dependencies, we developed an emulated physical layer (e-PHY). The e-PHY is integrated into \gls{oai} and replaces only the acquisition of PHY-layer samples. Instead of reading samples from an RF front end, each instance replays a previously recorded baseband trace in real time.  

The e-PHY is transparent to the upper layers. Each \gls{gnb} processes the replayed samples as if they had been acquired from a physical radio unit. At the same time, RUN-O-RAN receives the resulting information from the \gls{du} through the same E2 control and telemetry path. The xApp does not access the trace files directly; from its perspective, the three nodes behave as independent \glspl{gnb} observing the same uplink transmission through distinct propagation channels.

The three e-PHY instances replay three synchronized real traces \textit{DATA}$_m$, \textit{DATA}$_{s1}$, and \textit{DATA}$_{s2}$ according to the original transmission timing, thereby allowing the system to operate in real time.  
This approach allows the same signal observations to be replayed across multiple experiments. Different ranging, correction, and localization algorithms can therefore be compared using identical waveforms, hardware impairments, anchor geometries, and propagation conditions. 

For each scenario, we generate the channel realizations using the Sionna ray-tracing framework~\cite{sionna}. We convert each uplink path to .asc format and feed it to a Keysight PROPSIM channel emulator~\cite{keysight_propsim} with the propagation delay, attenuation, multipath components, and fading profile. 

The traces replayed by the e-PHY are collected using the testbed shown in Fig.~\ref{fig: siopsim}. The \gls{gnb} is an \gls{oai} instance attached to an Ettus Research USRP N310 software-defined radio~\cite{ettus_n310}, operating in band n77 with a $60~MHz$ channel bandwidth and numerology $\mu = 1$ ($30~kHz$ \gls{scs}). The radio front end is configured through PROPSIM with one \gls{dl} and three \gls{ul} paths. The \gls{dl} provides master-\gls{ue} connectivity, while the \glspl{ul} simultaneously capture the \gls{ue} \gls{srs} transmission after it has propagated through independently configured channels. The resulting baseband signals represent the observations available at the three anchors in a geographically distributed arbitrary deployment. 

For every \gls{srs} occasion, the setup records the three baseband traces. Because we used a COTS device to generate the waveforms (a Sierra Wireless~\cite{sierra_wireless} with a Qualcomm Snapdragon X60 5G Modem-RF system and a programmable SIM), the traces include oscillator behavior, transmission-timing adjustments, chipset-dependent effects, and other hardware imperfections.

The testbed combines commercial-device realism with controlled and repeatable propagation conditions, enabling performance evaluation under arbitrarily varying system parameters. Any experiment is fully determined by the recorded \gls{ue} waveforms. The same traces can therefore be reused across multiple runs, ensuring that all evaluated algorithms operate on the same uplink observations. 


%% file: Content/resultsTEST.tex
\section{Results}
\label{sc: results}

We evaluate RUN-O-RAN over an experimental dataset of $150{,}000$ \gls{srs} transmissions, using both \gls{nls} and \gls{ekf} estimators. The dataset includes open-square and narrow-street urban scenarios, under both static and dynamic \gls{ue} conditions. We selected these scenarios to stress the three factors that most directly affect uplink trilateration: anchor geometry, multipath-induced ranging bias, and \gls{srs} signal quality.

The static e-PHY dataset consists of eight measurement campaigns generated with the Sionna ray-tracing framework, using a detailed 3D model of the city of Munich. In the six static campaigns, the \gls{ue} is positioned in an open urban square, and the three anchors are placed on surrounding buildings at distances of 100\,m and 200\,m to evaluate the effect of multipath in \gls{los} and \gls{nlos}. Unless otherwise stated, the anchors are arranged at the vertices of an equilateral triangle centered on the \gls{ue}, so that the impact of propagation can be studied independently of \gls{gdop}. The last two campaigns then assess RUN-O-RAN under realistic dynamic scenarios, moving the \gls{ue} along a trajectory in the square and in a narrower street with an asymmetric anchor disposition. Fig.~\ref{fig: dynamic} depicts an example scenario.

This section aims not only to report localization accuracy, but also to identify where the error originates. We therefore organize the evaluation around four questions. \textit{(A)} How does the system behave under noisy \gls{srs}? \textit{(B)} Does RUN-O-RAN behave consistently with the geometry of trilateration? \textit{(C)} How does multipath affect the corrected range estimates and the final position? \textit{(D)} How do anchor-wise ranging errors propagate into localization error? 

\subsection{Noisy Ranging}
\label{sbs: results snr}

\begin{figure}[t]
\centering
\begin{subfigure}{\linewidth}
\includegraphics[width=\linewidth]{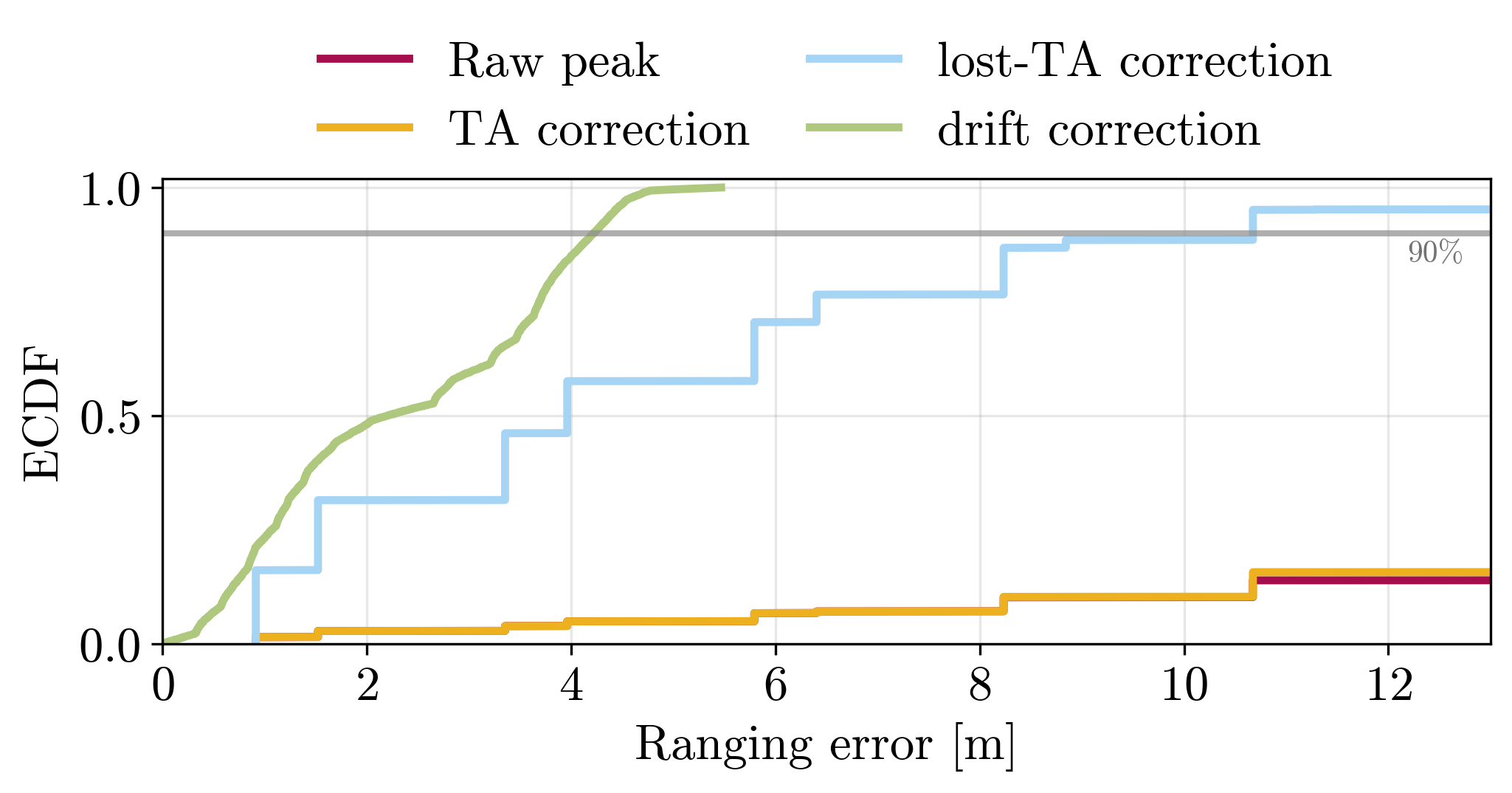} 
\caption{\footnotesize \acrshort{ecdf} of the ranging error after each correction step: raw peak detection, \acrshort{ta} correction, lost-\acrshort{ta} correction, and drift correction.}
\label{fig: ranging steps}
\end{subfigure}
\vfill
\begin{subfigure}{\linewidth}
\includegraphics[width=\linewidth]{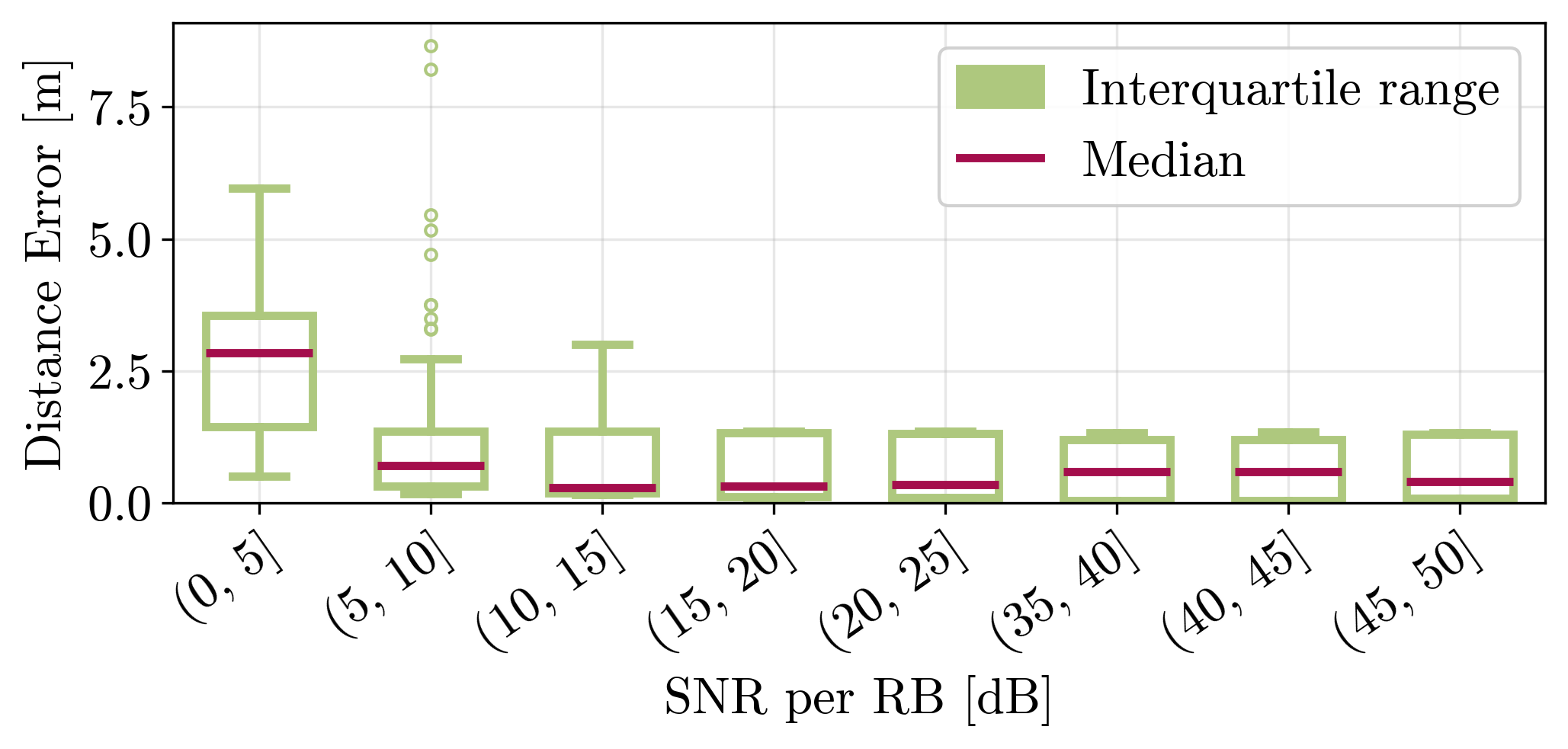} 
\caption{\footnotesize Impact of \acrshort{srs} \acrshort{snr} on ranging accuracy. Samples are grouped into $5$~dB \acrshort{snr} bins.}
\label{fig: SNR error}
\end{subfigure}
\caption{\footnotesize Impact of the correction pipeline and of \acrshort{srs} \acrshort{snr} on ranging accuracy in static conditions.}
\label{fig: noisy ranging}
\end{figure}

To give the reader an idea of the impact of the filters described in Sec.~\ref{sbs: positioning-pipeline}, we show in Fig.~\ref{fig: ranging steps} the \gls{ecdf} of the ranging error in static conditions after the sequence of correction steps: \gls{ta} correction, lost-\gls{ta} correction, and drift correction. The raw peak detection and \gls{ta}-corrected estimates are dominated by residual \gls{ta} missed updates, while the final drift-corrected estimate achieves a $90$th-percentile error of $4.22$~m, nearly an order of magnitude lower than the uncorrected case.

We then evaluate the relationship between the \gls{snr} of the received \gls{srs} and the ranging error. Lower \gls{snr} increases the uncertainty of the estimated \gls{cir} and makes the detection of the earliest significant peak less reliable. Unlike \gls{nlos} multipath, however, low \gls{snr} primarily increases measurement variance rather than introducing a persistent geometric bias.

Fig.~\ref{fig: SNR error} reports the ranging error of the previously analyzed scenarios, grouped into $5$~dB \gls{snr} bins. As expected, low-\gls{snr} samples exhibit larger dispersion, reflecting the increased difficulty of detecting the first arrival path in noisy channel conditions. Nevertheless, most estimates remain within a few meters as long as the \gls{ue} remains attached to its master.

This result indicates that \gls{snr} is not the primary source of error for positioning applications; however, an \gls{snr}-based rejection or reweighting rule may improve robustness against noisy samples.

\subsection{Impact of Anchor Geometry}
\label{sbs: results geometry}

\begin{figure}
\centering
\begin{subfigure}{\linewidth}
\centering
\begin{tikzpicture}[scale=0.7]
    
    \coordinate (gnbm) at (-2,0);
    \coordinate (gnbs1) at (6,0);
    \coordinate (gnbs2) at (2,3);
    
    \coordinate (ue) at (2,1);
        
    \fill (gnbm) circle (2pt);
    \fill (gnbs1) circle (2pt);
    \fill (gnbs2) circle (2pt);
    
    \node[below] at (gnbm) {\textit{gNB}$_m$};
    \node[below] at (gnbs1) {\textit{gNB}$_{s1}$};
    \node[above] at (gnbs2) {\textit{gNB}$_{s2}$};
    
    \draw[thick] (ue) ++(-0.12,-0.12) -- ++(0.24,0.24);
    \draw[thick] (ue) ++(-0.12,0.12) -- ++(0.24,-0.24);
    \node[above] at (ue) {UE};
    
    \draw[-, dashed] ($(gnbm)+(0.2,0.1)$) -- (1.8,0.9);
    \draw[-, dashed] ($(gnbs1)-(0.2,-0.1)$) -- (2.15,0.9);

    \draw[-] (gnbm) -- (gnbs1);
    \draw[-] (gnbs1) -- (gnbs2);
    \draw[-] (gnbm) -- (gnbs2);

    \pic[draw,thick, roxxo,angle radius=0.35cm] {angle = gnbm--ue--gnbs1};
    \node[below left, roxxo] at (1.5,0.8) {$\theta$};
        
    \end{tikzpicture}
\label{fig: geometry angle}
\end{subfigure} 
\vfill
\begin{subfigure}{\linewidth}
\centering
\includegraphics[width=\linewidth]{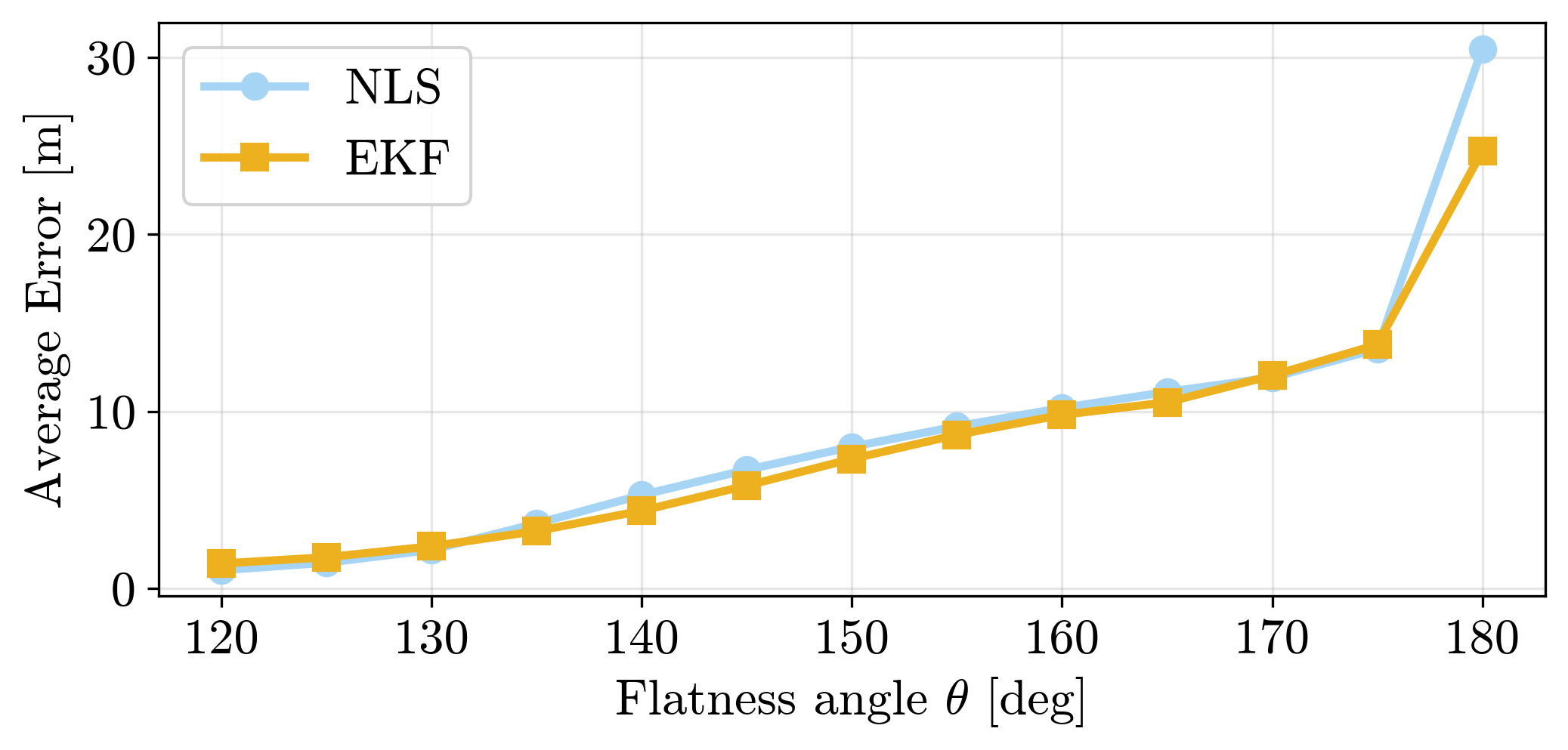}
\label{fig: geometry result}
\end{subfigure}
\vspace{-1cm}
\caption{\footnotesize Impact of anchor geometry on positioning error. The localization error increases as the anchor layout moves from a well-conditioned triangular configuration towards the degenerate aligned case.}
\label{fig: geometry gdop}
\end{figure}
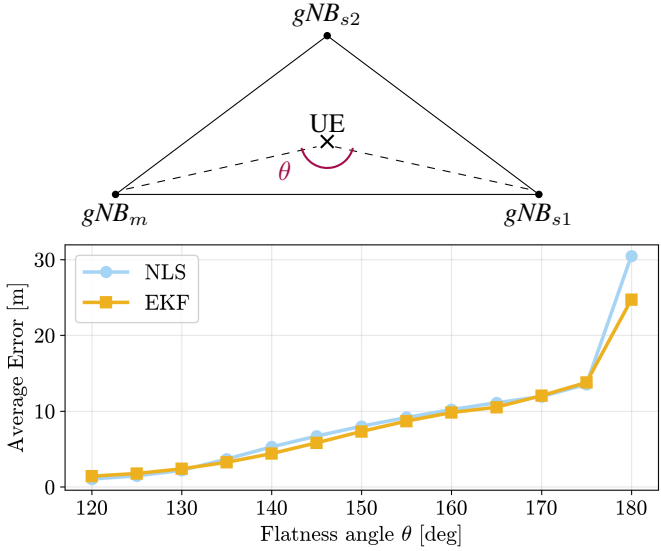

In this analysis, we aim to isolate the effect of anchor geometry; therefore, we conducted the experimental campaign in a square in Munich, with all anchors in the \gls{ue}'s \gls{los}.
In the scenario, the \gls{ue} is the centroid of a triangular configuration, where the centroid-vertex distance is fixed at $100$~m, shown in Fig.~\ref {fig: geometry gdop} above. The \glspl{gnb} are the vertices of the triangle and move along the circumference centered on the \gls{ue} and of radius $100$~m to change the geometry of the setup. We conducted the experiments by varying the widest angle at the centroid from $120^\circ$ to $180^\circ$ in $5^\circ$ increments. The $120^\circ$ case corresponds to an equilateral triangle, whereas the $180^\circ$ case corresponds to the degenerate configuration in which the anchors are aligned.

Fig.~\ref {fig: geometry gdop} on the left confirms the expected behavior of a trilateration system. Both the \gls{nls} and \gls{ekf} estimators produce comparable error distributions. Since the \gls{ue} is static, the \gls{ekf}'s temporal filtering provides little additional information over the memoryless \gls{nls} estimate. The estimators achieve their lowest error when the anchors surround the target with a well-conditioned geometry. As the triangle becomes increasingly obtuse, the same ranging uncertainty is amplified into a larger position error. The error peaks near the aligned configuration, where trilateration becomes ill-conditioned.

This result validates RUN-O-RAN's geometric consistency. The framework introduces no unexpected behavior: when the anchor layout degenerates, positioning error increases regardless of the estimator.

\subsection{Impact of Multipath}
\label{sbs: results multipath}
Multipath is a major radio impairment affecting \gls{srs}-based ranging in urban environments. We evaluate its impact on RUN-O-RAN under two propagation conditions, \gls{los} and \gls{nlos}. Channel realizations are generated with Sionna-rt in an urban square in Munich, with the \gls{ue} placed at the center and the three \glspl{gnb} at the vertices of an equilateral triangle centered on the \gls{ue}; this symmetric layout keeps the \gls{gdop} constant across scenarios, so that any difference in localization error can be attributed to multipath rather than to anchor geometry. The \glspl{gnb} are placed on the buildings surrounding the \gls{ue}, at a distance of $100$~m for the \gls{los} case and, to complete the study, at $200$~m for a \gls{nlos} feasibility campaign, where surrounding buildings obstruct the direct \gls{ue}-\gls{gnb} path.

We generate three levels of multipath richness: MP0, where only the direct path contributes to each \gls{ue}--\gls{gnb} channel; MP3, where we add up to three reflections; and MP5, where we add up to five reflections. We do not consider higher-order reflections because their peak amplitudes fall below the emulator's noise floor and are indistinguishable from zero.

\begin{figure}[h]
\centering
\begin{subfigure}{\linewidth}
\centering
\includegraphics[width=\linewidth]{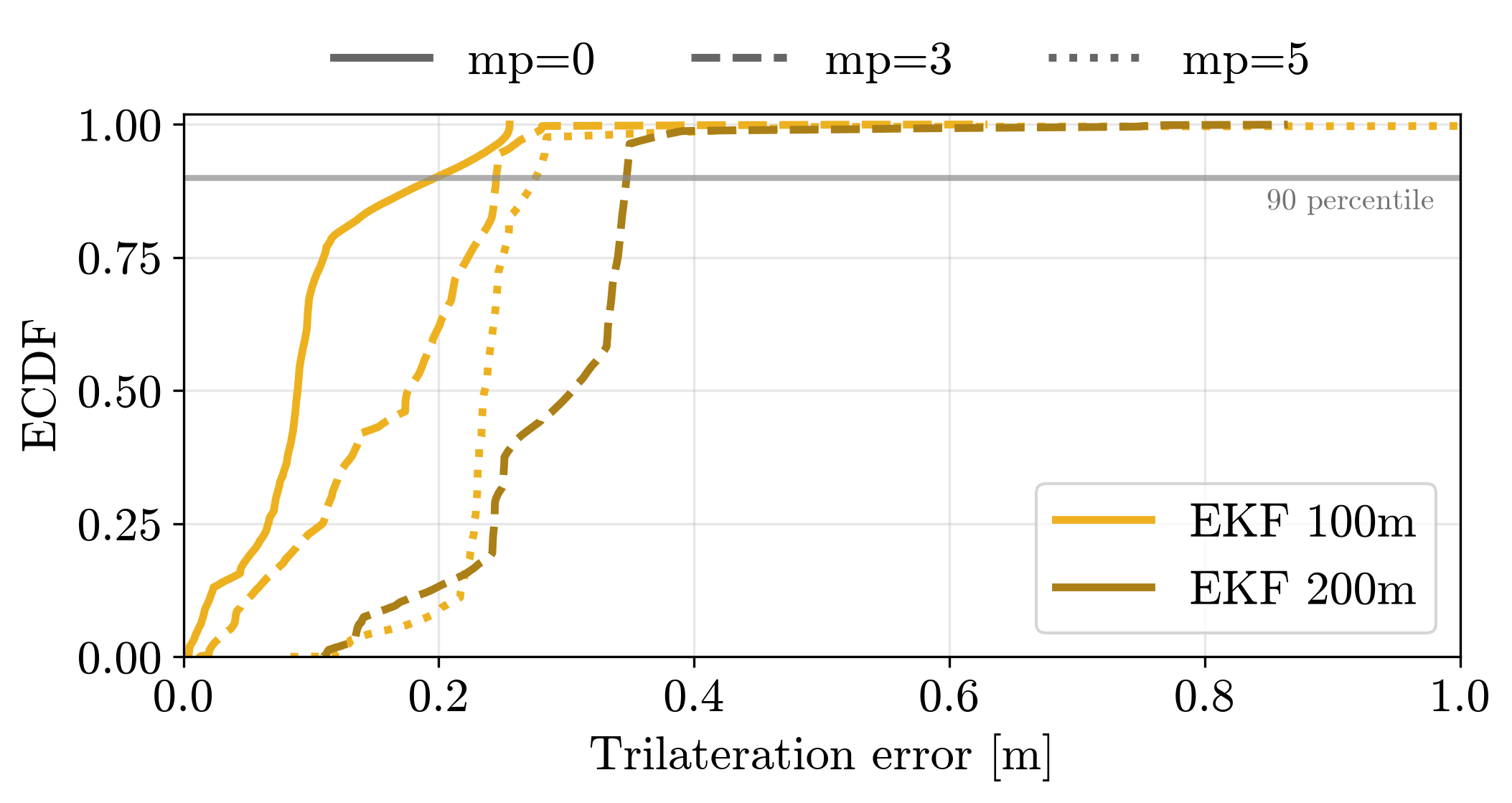}
\caption{\footnotesize EKF: LOS ($100$~m) vs. NLOS ($200$~m).}
\label{fig: ECDF EKF}
\end{subfigure}
\vfill
\begin{subfigure}{\linewidth}
\centering
\includegraphics[width=\linewidth]{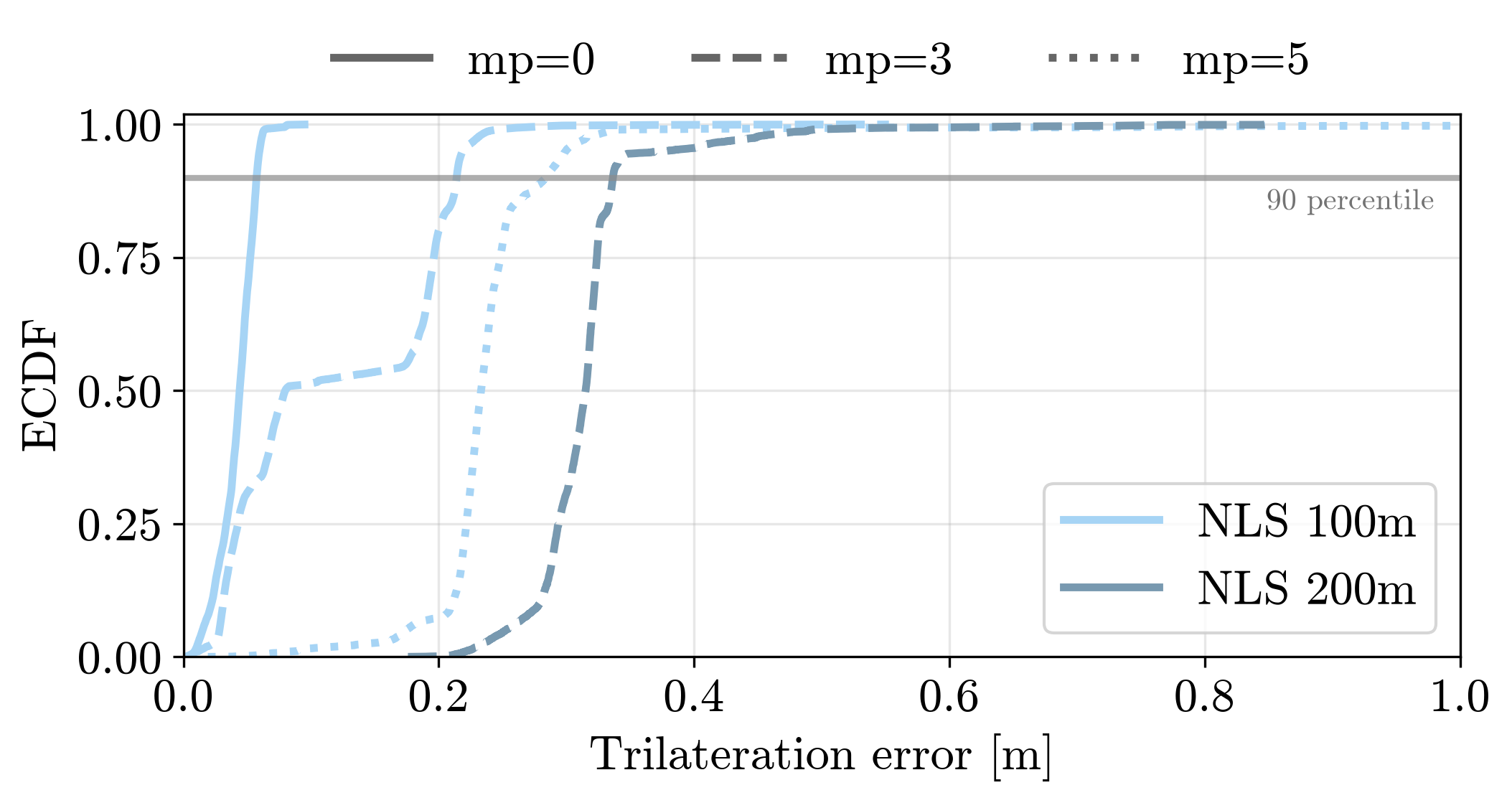}
\caption{\footnotesize NLS: LOS ($100$~m) vs. NLOS ($200$~m).}
\label{fig: ECDF NLS}
\end{subfigure} 
\caption{\footnotesize ECDF of the localization error under increasing multipath richness. MP0, MP3, and MP5 denote scenarios with zero, three, and five reflections, respectively.}
\label{fig: ECDF}
\end{figure}

Fig.~\ref{fig: ECDF} reports the \gls{ecdf} of the localization error for the \gls{ekf} (Fig.~\ref{fig: ECDF EKF}) and \gls{nls} (Fig.~\ref{fig: ECDF NLS}) estimators, comparing the three multipath levels in \gls{los} ($100$~m) against \gls{nlos} ($200$~m) scenario. 

Both Fig.~\ref{fig: ECDF EKF} and Fig.~\ref{fig: ECDF NLS} show that, already at $100$~m, increasing multipath richness clearly degrades the position estimate: the $90$-th percentile grows from a few centimeters at MP0 to several tens of centimeters at MP5. As expected, the \gls{nlos} campaign yields the worst \gls{ecdf} among all the configurations considered. This follows from how RUN-O-RAN mitigates multipath: it selects the earliest significant peak of the estimated \gls{cir}, rather than the strongest, assuming the first arrival is the direct path. In \gls{los}, this assumption holds, and the range estimate closely matches the true distance. In \gls{nlos}, the direct path is blocked, so even the earliest detected component is a reflection. This residual excess length manifests as a positive bias in the range estimate and, consequently, in the localization error.

\subsection{Propagation of Ranging Errors}
\label{sbs: results ranging-propagation}

We investigate how anchor-wise ranging errors propagate into the final position estimate. We restrict this analysis to the \gls{nls} estimator, whose output depends solely on the current set of range measurements. In contrast, the \gls{ekf} estimate is also conditioned on previous filter states, which combines the contribution of the current measurement with that of the estimator's memory and precludes a clean attribution of error to a specific anchor.

\begin{figure}[t]
\centering
\includegraphics[width=0.8\linewidth]{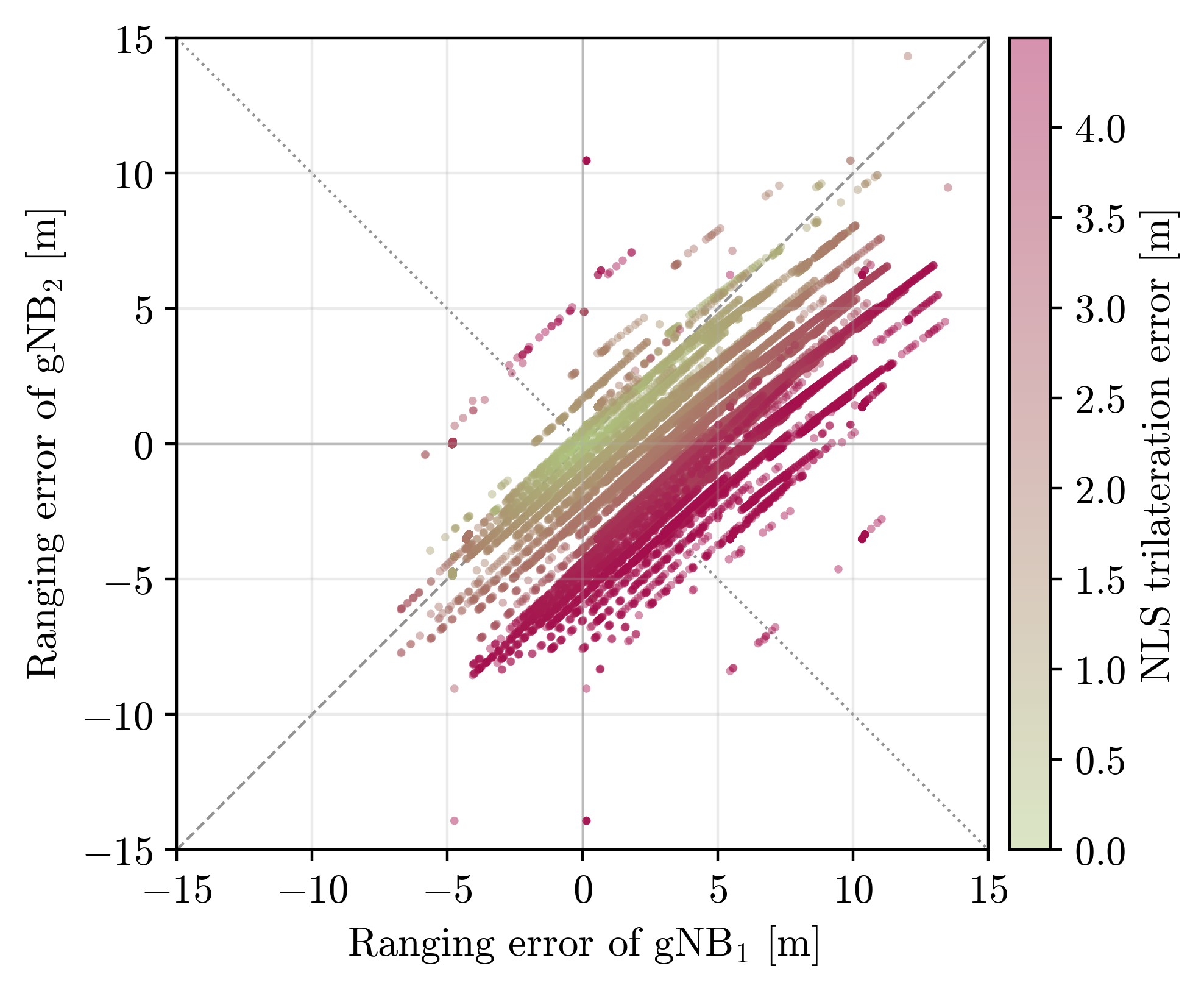} 
\caption{\footnotesize Propagation of anchor-wise ranging errors into the NLS positioning error. Each axis represents the ranging error associated with one gNB; color intensity encodes the resulting localization error.}
\label{fig: accuracy of ranging}
\end{figure}

We characterize this propagation using all our experimental campaigns. To isolate the secondary anchors' contribution to the trilateration outcome, we built a controlled post-processing dataset: we retain RUN-O-RAN's range estimates for both secondary gNBs, while replacing the master's range with its ground-truth value. This construction also supports per-anchor analysis, isolating cases where only one secondary is significantly biased from cases where both are.

Fig.~\ref{fig: accuracy of ranging} maps the ranging error of each secondary anchor onto the resulting localization error. Each axis corresponds to one secondary anchor's ranging error, while the color of each scattered point encodes the associated localization error. The scatter is denser below the $y = x$ diagonal, indicating that gNB\textsubscript{1}'s ranging error typically exceeds gNB\textsubscript{2}'s. Along the diagonal, where both anchors are biased by comparable amounts, the localization error remains limited: same-sign errors largely cancel in the trilateration solution, leaving the estimated position close to the true \gls{ue} location. Moving toward the anti-diagonal, where the two errors carry opposite signs, the localization error grows sharply, as the equilateral anchor geometry amplifies the mismatch and displaces the estimate well beyond the true one.

This representation makes clear that localization degradation is not uniformly distributed across anchors: a single biased \gls{gnb} can dominate the resulting position error, while bias on both anchors compounds it. Beyond characterizing this asymmetry, the plot is diagnostic in practice, as it can identify which anchor contributes most to the localization error in a given deployment.

\subsection{Dynamic Scenarios}
\label{sbs: results dynamic}

\begin{figure}[t]
\centering
\hspace{0.56 cm}
\begin{subfigure}{0.41\linewidth}
\centering
\includegraphics[trim={0 90 0 0}, clip, width=\linewidth]{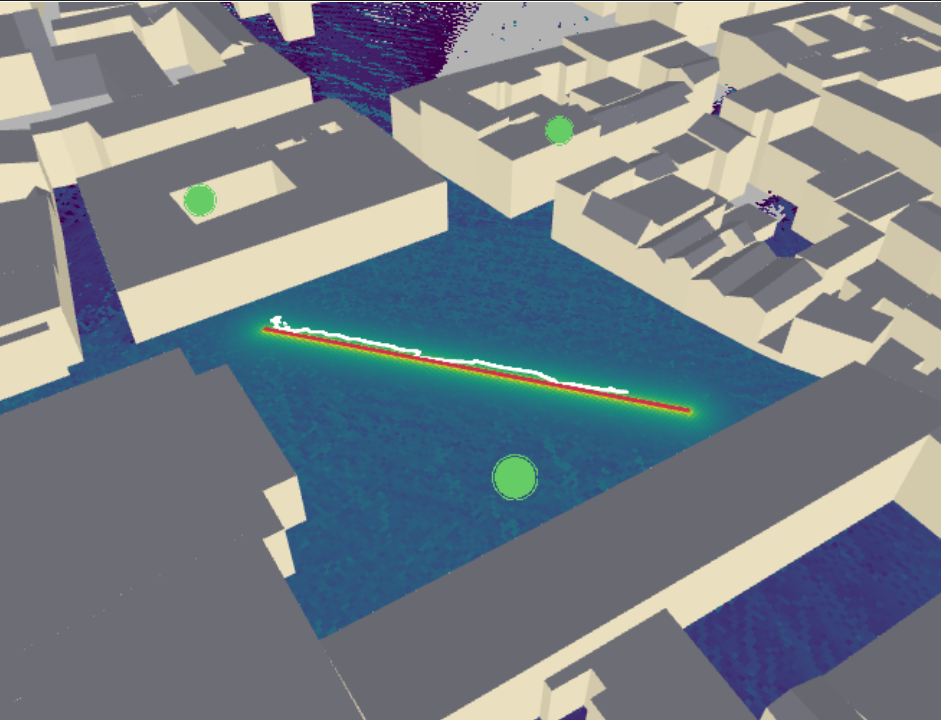}
\caption{\footnotesize Square scene.}
\label{fig: squareline}
\end{subfigure}
\hspace{0.1 cm}
\begin{subfigure}{0.41\linewidth}
\centering
\includegraphics[width=\linewidth]{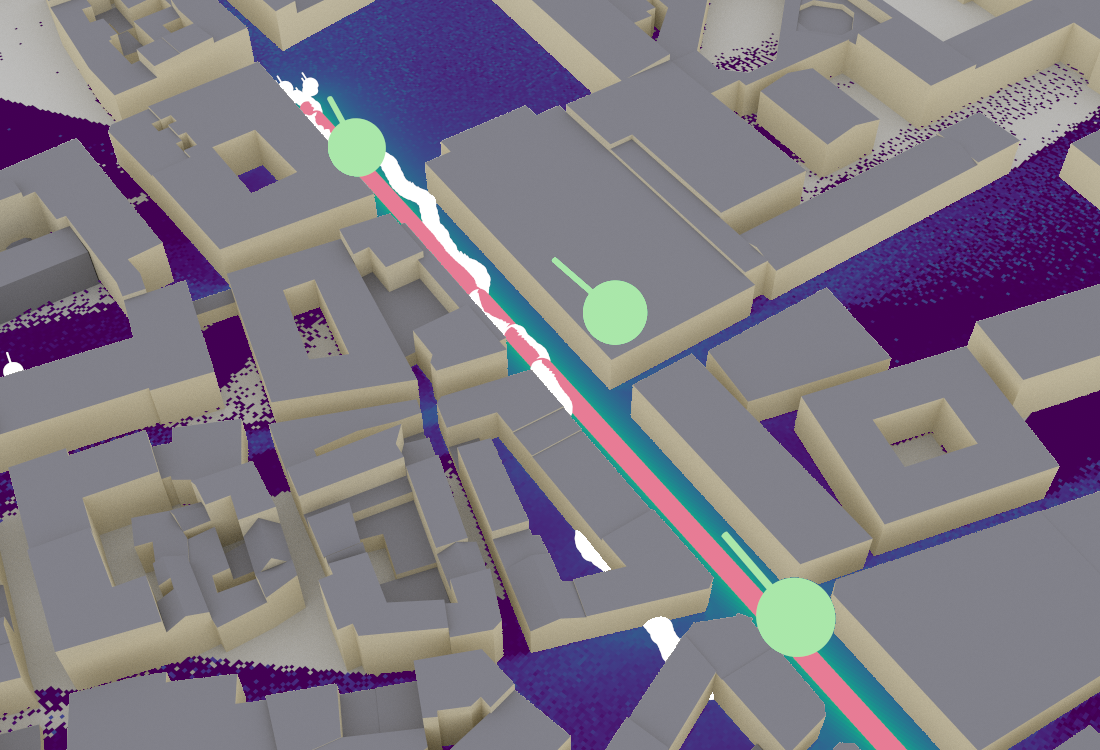}
\caption{\footnotesize Street scene.}
\label{fig: street}
\end{subfigure}
\bigskip
\begin{subfigure}{\linewidth}
\centering
\includegraphics[width=\linewidth]{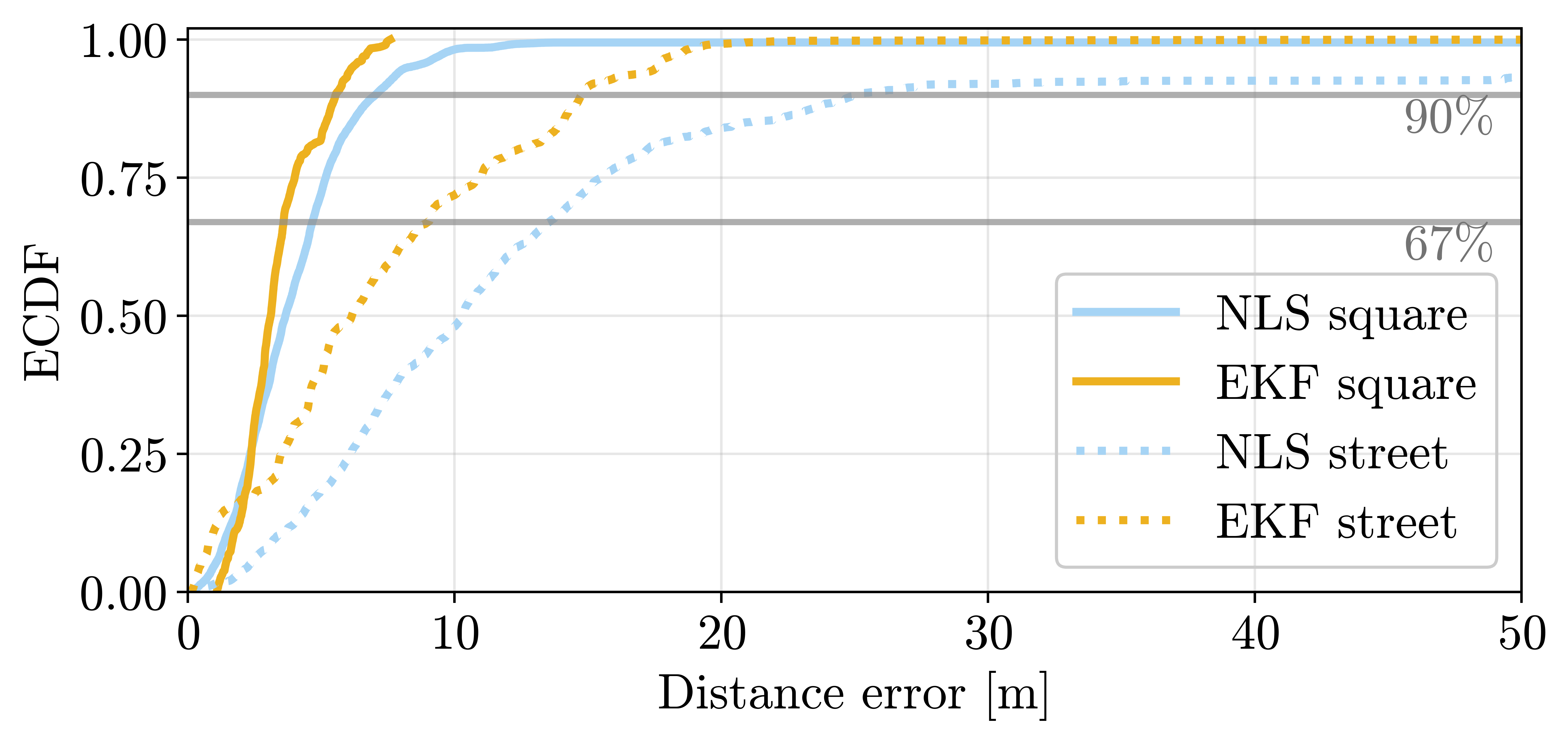}
\caption{\footnotesize ECDF.}
\label{fig: ECDF dynamic}
\end{subfigure}
\vspace{-1cm}
\caption{\footnotesize Dynamic scenarios: square and street. Fig.~\ref{fig: squareline}--~\ref{fig: street} show in red the ground truth, in white the estimated trace, and in green the three anchors. Fig.~\ref{fig: ECDF dynamic} shows the localization error for the dynamic scenarios. Horizontal lines mark the 67th and 90th percentiles.}
\label{fig: dynamic}
\end{figure}

Finally, we evaluate RUN-O-RAN under \gls{ue} mobility. The dynamic dataset includes two pedestrian-like trajectories in the Munich environment: a \gls{ue} crossing an open square, and a \gls{ue} moving through a narrow urban canyon. The open-square scenarios preserve a favorable anchor geometry, whereas the urban-canyon scenario combines stronger multipath with a less favorable layout.

As expected, in dynamic conditions, the difference between the two estimators becomes more visible. The \gls{nls} estimator processes each \gls{srs} instant independently and therefore reflects the instantaneous quality of the current range estimates. The \gls{ekf} uses the same geometric measurement model but adds a constant-velocity prior, thereby suppressing isolated outliers and producing smoother trajectories.

The experimental results confirm this behavior. Fig.~\ref{fig: ECDF dynamic} shows the resulting ECDFs for both estimators and both scenes: in open-square trajectories, where the geometry is favorable and \gls{los} propagation is more likely, both estimators provide stable localization, with the \gls{ekf} reducing abrupt fluctuations caused by transient ranging errors. In the urban-canyon trajectory, performance degrades because \gls{nlos} propagation introduces persistent range bias, and the corresponding curves in Fig.~\ref{fig: ECDF dynamic} are shifted well to the right of the square-scene curves. In this case, the \gls{ekf} smooths the trajectory but cannot fully remove the systematic displacement caused by biased measurements.

This distinction is essential. Temporal filtering improves stability when errors are sporadic, but it cannot compensate for persistent \gls{nlos} bias without additional information. Therefore, the dynamic experiments confirm the feasibility of real-time RUN-O-RAN tracking while highlighting a key direction for further improvements: detecting and mitigating \gls{nlos}-induced range errors through anchor selection or propagation-aware models based on environment digital twins.




%% file: Content/discussion.tex
\section{Discussion: Synchronization and Anchor Reliability}
\label{sc: discussion}

RUN-O-RAN demonstrates that cooperative uplink localization can be implemented as an O-RAN-native service. Building on these results, we discuss two aspects that are key to extending its applicability to broader deployment scenarios: inter-\gls{gnb} synchronization and multi-anchor availability.
\subsection{Inter-gNB Synchronization}
\label{sbs: discussion synchronization}

Operational cellular networks already rely on synchronization mechanisms such as \gls{gnss}-based timing, SyncE, and IEEE~1588 PTP, especially in TDD deployments~\cite{ericsson_sync, ericsson_sync_key, microchip_sync}. For localization, however, residual relative timing error among the anchors can introduce noise into the estimates. In this section, we highlight the impact of three different synchronization errors across \glspl{gnb}: 1) timing offset, 2) phase noise, and 3) frequency drift.
\vspace{0.3cm}
\subsubsection{Timing Offset}
A constant bias $b_{i,m}$ between the local clocks of anchor $i$ and the master $m$ shifts the estimated differential delay by a fixed, time-invariant amount:
\begin{equation}
\Delta\hat{\tau}_{i,m}(t) = \Delta\tau_{i,m}(t) + b_{i,m}.
\end{equation}%
Since $b_{i,m}$ does not evolve over the observation window, it is therefore fully absorbed by the existing calibration procedure for hardware asymmetries~\ref{sbbs: xapp-calibration} and requires no dedicated compensation logic.
\vspace{0.3cm}
\subsubsection{Phase Noise}
Residual oscillator phase noise introduces a random, zero-mean fluctuation to the distance estimation:
\begin{equation}
    \phi_{i,m}(t)\sim \left(\mathbb{E}[\phi_{i,m}(t)] = 0, \sigma_{\phi_{i,m}}^2 = \mathbb{E}[\phi_{i,m}^2(t)]\right).
\end{equation}%
The standard deviation $\sigma_{\phi_{i,m}}$ is set to $33$\,m; this value represents an upper bound on the phase-noise-induced ranging error observed in real 5G deployments~\cite{ClockBias}.

\begin{figure}[t]
    \centering
    \begin{subfigure}{\linewidth}
    \includegraphics[width=0.9\linewidth]{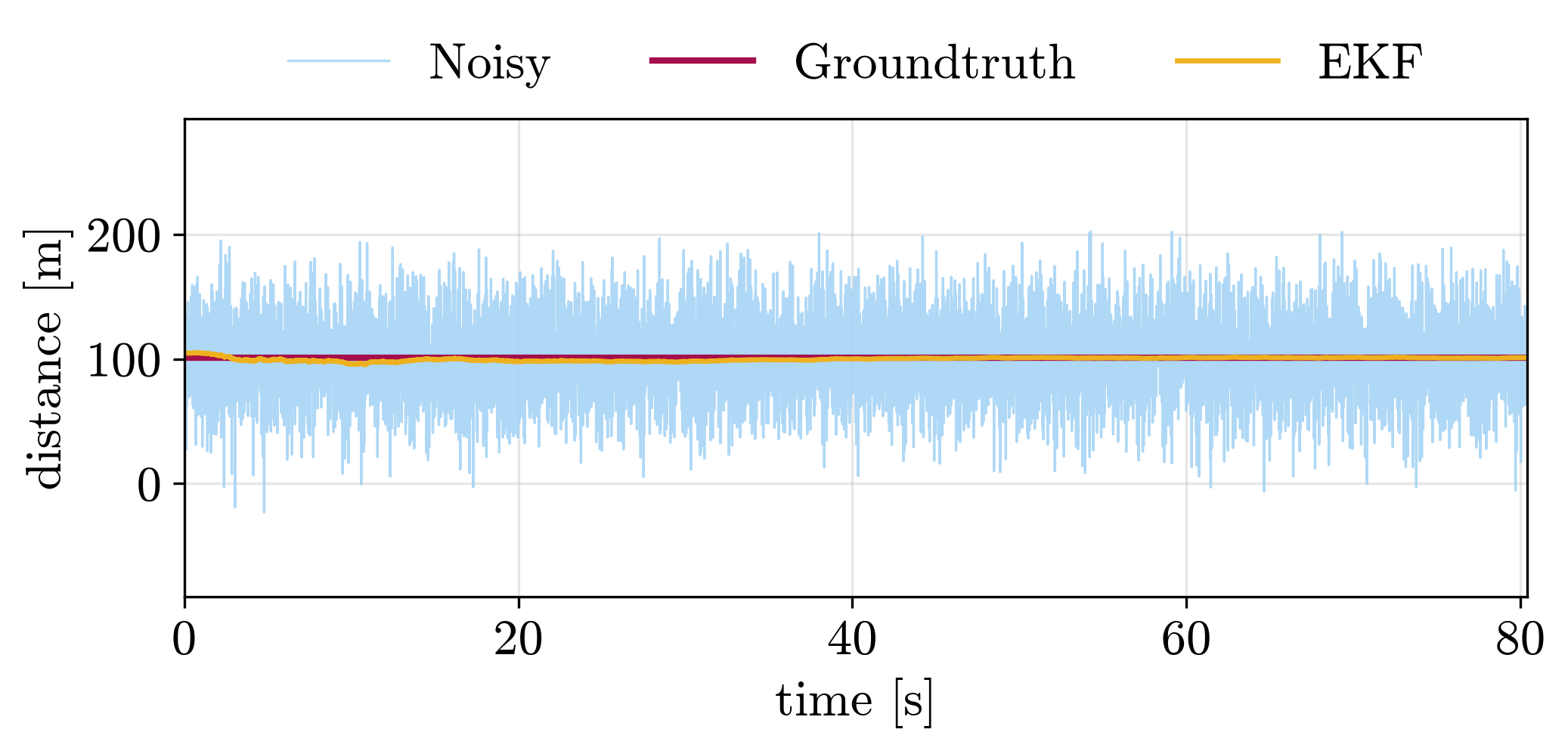}
    \caption{\footnotesize EKF phase noise correction.}
    \label{fig: phase_noise_EKF}
    \end{subfigure}
    \vfill
    \begin{subfigure}{\linewidth}
    \includegraphics[width=0.9\linewidth]{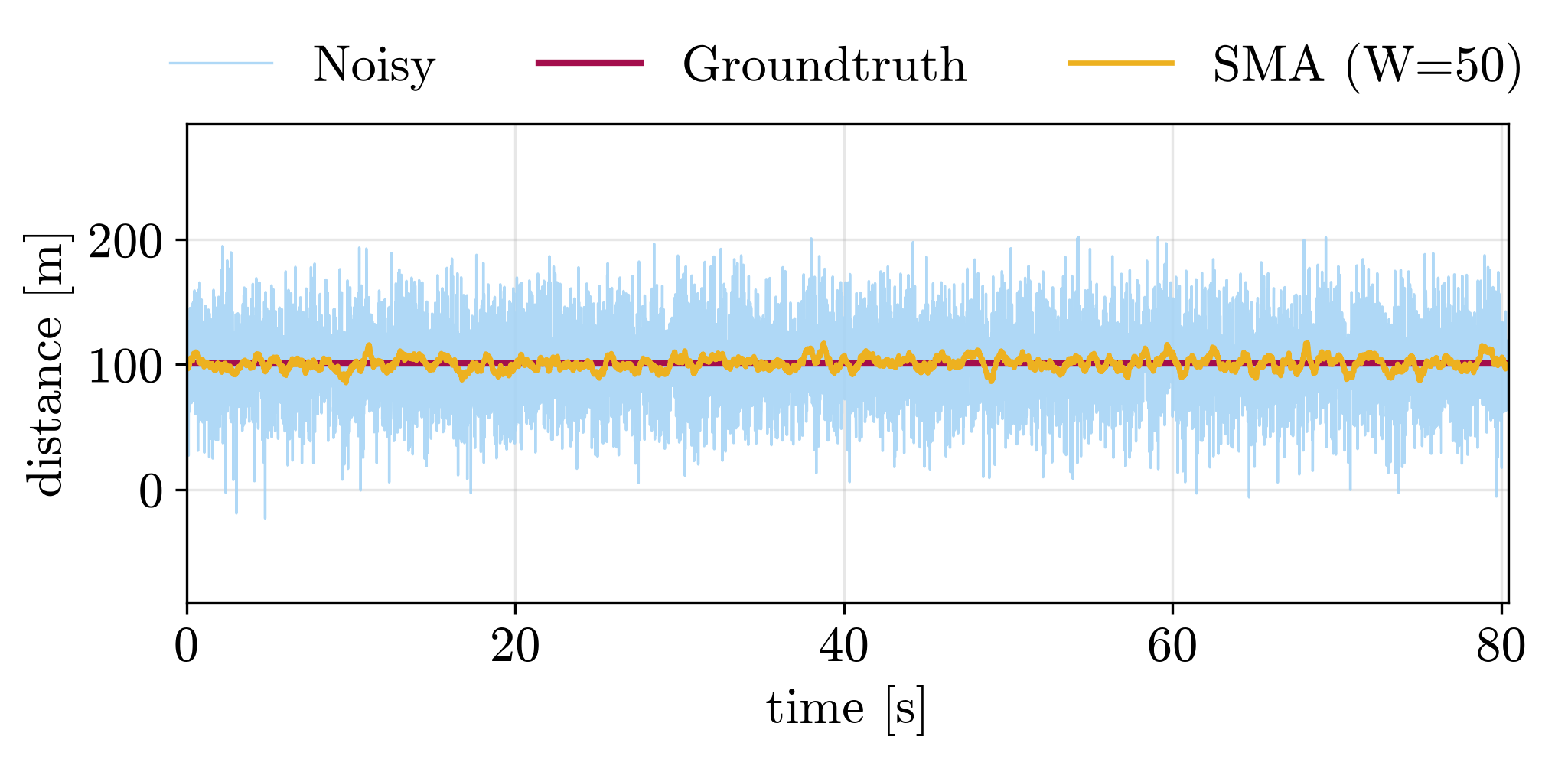}
    \caption{\footnotesize SMA phase noise correction.}
    \label{fig: phase_noise_W}
    \end{subfigure}
    \caption{\footnotesize Phase noise affecting \gls{gnb}$_s$ in ranging estimation, corrected through EKF and moving average.}
    \label{fig: phase_noise}
\end{figure}

This error is stationary and, to first approximation, uncorrelated across measurement epochs, so it does not accumulate over time. Its contribution to the range estimate can be reduced by applying the \gls{ekf} or by applying a Simple Moving Average (SMA) over $W$ independent samples, which reduces the impact to $\nicefrac{\sigma_{\phi_{i,m}}}{\sqrt{W}}$. Fig.~\ref{fig: phase_noise} shows the estimated ranging distance for a secondary anchor $i$ with respect to the master $m$: the raw ranging (light blue) fluctuates around the ground truth (red), while both SMA and \gls{ekf} (orange) suppress this fluctuation, with the \gls{ekf} tracking the ground truth more tightly.
\vspace{0.3cm}
\subsubsection{Frequency Drift}
Clock drift is the most critical because it introduces a time-varying error that accumulates across consecutive measurements. The current implementation compensates for timing impairments associated with the \gls{ue} and the serving-cell timing process. Extending the same principle to inter-\gls{gnb} drift is a natural continuation. For a static \gls{ue}, the corrected differential delay between two anchors should remain constant once propagation and \gls{ta} effects have been removed. A systematic variation observed in the slave anchors therefore indicates residual relative clock drift:
\begin{equation}
\hat{\delta}_{i,m}
=
\frac{d}{dt}\Delta \hat{\tau}_{i,m}(t),
\end{equation}
where $\Delta \hat{\tau}_{i,j}(t)$ is the differential delay between anchors $i$ and the master. In ideal conditions, $\hat{\delta}_{i,m}=0$. When this is not the case, the same drift-estimation logic used in Sec.~\ref{sbbs: second-drift} can be applied to estimate $\hat{\delta}_{i,j}$ and compensate for the corresponding accumulating range error. In this case, Eq.~\ref{eq: driftone} would become:
\begin{equation}
\label{eq: driftone}
\tilde{d}_{\mathrm{UE},i}
= \hat{d}_{\mathrm{UE},i}^{\, \mathrm{TA\text{-}Corr.}}
- c \hat{\delta}_{\nicefrac{+}{-}} \Delta t_{\mathrm{sync}}
- c \hat{\delta}_{i,m} \Delta t_{i,m\,\mathrm{sync}}.
\end{equation}
This extension is beyond the scope of the present evaluation, but RUN-O-RAN enables it directly, as the xApp observes the same uplink transmission from multiple anchors.

\subsection{Multi-Cell Visibility and Anchor Availability}
\label{sbs: discussion coverage}

RUN-O-RAN assumes that the target \gls{ue} is observable by at least three anchors with compatible radio configurations. This condition is plausible in dense small-cell deployments, private 5G networks, industrial campuses, and indoor distributed deployments, but it is not guaranteed in sparse or frequency-fragmented networks. In those cases, fewer than three \glspl{gnb} may be able to receive and process the same \gls{srs} transmission.

This limitation defines the operating region of pure trilateration, but it does not limit the broader RUN-O-RAN architecture. When three anchors are unavailable, the framework can be extended by combining range measurements with angular or beam measurements. A multi-antenna \gls{gnb}, for example, can in principle fuse \gls{toa}-based ranging with \gls{aoa} estimation~\cite{albertoAoA}; with two anchors, range and angle constraints can reduce the remaining ambiguity. This points toward hybrid \gls{toa}/\gls{aoa} localization within the same O-RAN control framework.

More generally, RUN-O-RAN should be interpreted as a programmable infrastructure for network-side localization, rather than as a fixed three-anchor trilateration pipeline. Future xApps can select the best available subset of anchors according to geometry, \gls{snr}, \gls{nlos} presence, multipath likelihood, antenna configuration, resource availability, and synchronization quality~\cite{Xhafa}. 
This would let the localization service adapt to the deployment instead of assuming every scenario provides the same anchor set.